\documentclass[preprint,superscriptaddress,amsmath,amssymb,aps,eqsecnum,nofootinbib]{revtex4}
\usepackage[dvipdfmx]{graphicx}
\usepackage{subfigure}
\usepackage{bm}
\usepackage{mathrsfs}

\usepackage{color}
\usepackage{ulem}

\begin{document}
\title{Ladder operators and quasinormal modes in Ba\~{n}ados-Teitelboim-Zanelli black holes}
\author{Takuya Katagiri}
\affiliation{Astronomical Institute, Graduate School of Science, Tohoku University, Aoba, Sendai 980-8578, Japan}
\affiliation{Niels Bohr International Academy, Niels Bohr Institute, Blegdamsvej 17, 2100 Copenhagen, Denmark}
\author{Masashi Kimura}
\affiliation{Department of Informatics and Electronics, Daiichi Institute of Technology, Tokyo 110-0005, Japan}
\affiliation{Department of Physics, Rikkyo University, Toshima, Tokyo 171-8501, Japan}
\date{\today}

\begin{abstract}
We study quasinormal modes (QNMs) of massive Klein-Gordon fields in static Ba\~{n}ados-Teitelboim-Zanelli (BTZ) black holes in terms of ladder operators constructed from spacetime conformal symmetries.
Because the BTZ spacetime is locally isometric to the three-dimensional anti-de Sitter spacetime, 
ladder operators, which map a solution of the massive Klein-Gordon equation into 
that with different mass squared, can be constructed from spacetime conformal symmetries.
In this paper, we apply the ladder operators to the QNMs of the Klein-Gordon equations in the BTZ spacetime.
We demonstrate that the ladder operators can change indices of QNM overtones,
and all overtone modes can be generated from a fundamental mode when we impose the Dirichlet or Neumann boundary condition at infinity.
We also discuss the case with the Robin boundary condition.
\end{abstract}

\maketitle

\section{Introduction}
Ladder operators are useful tools that provide a deep insight into a system. For example, in quantum mechanics, they change a quantum number of solutions of the Schr\"odinger equation and allow for relating the different eigenstates without detailed knowledge of the solutions. It is known that the ladder operators in quantum mechanics are related to the underlying symmetry of a given system. 
As shown in~\cite{Cardoso:2017qmj,Cardoso:2017egd}, a similar discussion based on ladder operators from  symmetry of spacetimes is possible for a Klein-Gordon field. Ladder operators of the massive Klein-Gordon field can be defined in spacetimes with a particular conformal symmetry, e.g., (anti-)de Sitter spacetimes, and these operators change the mass squared of the Klein-Gordon field~\cite{Cardoso:2017qmj,Cardoso:2017egd}.
The operator is named the mass ladder operator and allows one to analyze the deeper structure of test fields in curved spacetimes~\cite{Katagiri:2021scx,Katagiri:2021xig} as ladder operators in quantum mechanics.

Test fields in a curved spacetime tell us a fundamental geometrical property of the spacetime. They play an important role in the understanding of phenomena in the strong-gravity regime, say, energy extraction phenomena from black holes~\cite{Brito:2015oca, Blandford:1977ds}, weak tidal fields in a compact binary system~\cite{Hinderer:2007mb,Binnington:2009bb}, ringdown gravitational waves~\cite{Vishveshwara:1970zz, Giesler:2019uxc}, and probe of strong cosmic censorship conjecture~\cite{Cardoso:2017soq}. In general, to obtain a physically meaningful solution of the equation of motion,
we need to impose appropriate boundary conditions on the solution. For example, for a test field in black hole spacetimes, appropriate boundary conditions at the horizon and infinity define its characteristic oscillation, namely, quasinormal ringing, which plays an important role in various contexts~\cite{Regge:1957td,Berti:2005ys,Hod:1998vk,Horowitz:1999jd,Cardoso:2001hn,Birmingham:2001pj,Kodama:2003jz,Ishibashi:2003ap,Kodama:2003kk,Berti:2009kk,Isi:2019aib}. 
Naturally, the following question arises: Does a solution generated from a physically meaningful solution by the mass ladder operator~\cite{Cardoso:2017qmj,Cardoso:2017egd} satisfy the appropriate boundary conditions that are the same as the original solution?

In this paper, we apply the mass ladder operators to phenomena, which highly depend on the global structure of the solution, namely, quasinormal modes (QNMs) of the massive Klein-Gordon field in static Ba\~{n}ados-Teitelboim-Zanelli~(BTZ) spacetimes~\cite{Banados:1992wn}.\footnote{In the previous work~\cite{Katagiri:2021xig}, we have mainly focused on phenomena, which depend only on the local property of the solution.} This setup is the simplest system, in which one can obtain an exact solution for QNMs and can globally define the mass ladder operators. With the rotational symmetry of the spacetime, the QNMs can be decomposed into independent angular modes with an integer~$m$. For each mode with $m$, there exists a discrete set of modes labeled by a nonnegative integer~$n$, which is an index of overtones. We particularly investigate, when acting the mass ladder operators on the QNMs, whether the operators generate QNMs with different indices~$(m,n)$. 

This paper is organized as follows. In Sec.~\ref{Sec:QNM}, we review QNMs and QNM frequencies of the massive Klein-Gordon field in the static BTZ spacetime. In Sec.~\ref{Sec:MLopandBCs}, we construct the mass ladder operators in the BTZ spacetime. In Sec.~\ref{Sec:MassladderandQNMBC}, we discuss the relation between the mass ladder operators and the QNM boundary conditions. In Sec.~\ref{Sec:MLopandQNMs}, we argue the shift of the QNM frequencies by the mass ladder operators. In Sec.~\ref{Sec: Remark}, we provide some remarks on our results. In Sec.~\ref{Sec:Summary}, we summarize this work. 

\section{Quasinormal modes in static BTZ spacetimes}
\label{Sec:QNM}
In this section, we review QNMs of the massive Klein-Gordon field in the static BTZ spacetime by following~\cite{Cardoso:2001hn,Birmingham:2001pj,Ichinose:1994rg}.

\subsection{Massive Klein-Gordon field in the BTZ spacetime}
In static coordinates~$\left(t,r,\varphi\right)$, the static BTZ spacetime is described by~\cite{Banados:1992wn}
\begin{equation}
\label{BTZ}
ds^2=-N^2(r)dt^2+\frac{1}{N^2(r)}dr^2+r^2d\varphi^2,~~N^2(r)=-M+\frac{r^2}{\ell^2},
\end{equation}
where we assume~$M>0$ and~$\ell$ is the length scale of three-dimensional anti-de Sitter spacetimes. The range of the coordinate~$\varphi$ is $0\le \varphi<2\pi$ and $r=r_H:=\ell\sqrt{M}$ is the horizon radius such that $N^2(r_H)=0$. We consider the massive Klein-Gordon field~$\Phi\left(t,r,\varphi\right)$ that satisfies
\begin{equation}
\label{EOM}
\left[\nabla_{\mu}\nabla^{\mu}-\mu^2\right]\Phi=0,
\end{equation}
where the mass squared satisfies $\mu^2\ge\mu^2_{\rm BF}$ 
and $\mu^2_{\rm BF}:=-1/\ell^2$ is the Breitenlohner-Freedman bound in three-dimensional anti-de Sitter spacetimes~\cite{Breitenlohner:1982jf,Breitenlohner:1982bm}.\footnote{In this  paper, we focus on the case $\mu^2 \ge \mu^2_{\rm BF}$.
This is because the system with $\mu^2 < \mu^2_{\rm BF}$ is generically unstable~\cite{Breitenlohner:1982jf,Breitenlohner:1982bm}.
Another reason is that the mass ladder operator, which is the main subject of the present paper, for the real-valued mass squared exists only for the case $\mu^2 \ge \mu^2_{\rm BF}$~\cite{Cardoso:2017qmj}.
} Expanding the field as
\begin{equation}
\begin{split}
\Phi\left(t,r,\varphi\right)=&\sum_m \phi_m(r)e^{-i\omega_m t}e^{im\varphi},
\end{split}
\end{equation}
where~$\omega_m\in\mathbb{C}$ and~$m\in\mathbb{Z}$, Eq.~\eqref{EOM} is reduced to an equation,
\begin{equation}
\label{eqpsi1}
\phi_{m}''+\left(\frac{1}{r}+\frac{\left(N^2\right)'}{N^2}\right)\phi_{m}'+\frac{1}{N^2}\left(\frac{{\omega_m}^2}{N^2}-\frac{m^2}{r^2}-\mu^2\right)\phi_{m}=0,
\end{equation}
where the prime denotes the derivative with respect to~$r$. Hereafter, we will omit the subscript~$m$ of~$\phi_m$ and~$\omega_m$. 

Equation~\eqref{eqpsi1} can be solved in terms of the hypergeometric functions.
We introduce a valuable~$z\in[0,1)$ defined by
\begin{equation}
z=1-\frac{{r_H}^2}{r^2},
\end{equation}
and a function~$f(z)$ such that
\begin{equation}
\phi(z)=z^{-i\frac{\omega\ell^2}{2r_H}}\left(1-z\right)^{\frac{1+\sqrt{1+\mu^2\ell^2}}{2}}f(z).
\end{equation}
Then, Eq.~\eqref{eqpsi1} can be written in the form,
\begin{equation}
\label{hypeq}
z\left(1-z\right)\frac{d^2}{dz^2} f+\left[c-\left(1+a+b\right)z\right]\frac{d}{dz}f-ab f=0,
\end{equation}
where constants $a, b, c$ are given by
\begin{equation}
\begin{split}
\label{abc}
a&=\frac{1+\sqrt{1+\mu^2\ell^2}}{2}-i\frac{\ell}{2r_H}\left(\omega\ell-m\right),\\
b&=\frac{1+\sqrt{1+\mu^2\ell^2}}{2}-i\frac{\ell}{2r_H}\left(\omega\ell+m\right),\\
c&=1-i\frac{\omega\ell^2}{r_H}.
\end{split}
\end{equation}
Equation~\eqref{hypeq} is the differential equation for Gaussian hypergeometric functions, of which linearly independent solutions are 
$\!_2F_1\left(a,b;c;z\right)$ and~$z^{1-c}~_2F_1\left(a-c+1,b-c+1;2-c;z\right)$. We thus obtain a general solution of Eq.~\eqref{eqpsi1}:
\begin{equation}
\begin{split}
\label{generalsols}
\phi(z)=z^{-i\frac{\omega\ell^2}{2r_H}}\left(1-z\right)^{\frac{1+\sqrt{1+\mu^2\ell^2}}{2}}&\left[A~_2F_1\left(a,b;c;z\right)\right.\\
&\left.~+Bz^{1-c}~_2F_1\left(a-c+1,b-c+1;2-c;z\right)\right],
\end{split}
\end{equation}
where~$A,B$ are constants. 

\subsection{QNM boundary condition at the black hole horizon}
To define QNMs, we impose appropriate boundary conditions on the general solution~\eqref{generalsols} at the black hole horizon and infinity. Near the horizon, the general solution behaves as
\begin{equation}
\phi \simeq  2^{-\frac{i\omega\ell^2}{2r_H}}A e^{-i\omega r_*} + 2^{\frac{i\omega\ell^2}{2r_H}} B e^{i\omega r_*}, 
\end{equation}
where $r_*$ is the tortoise coordinate defined by $dr_* = N^{-2}dr$ and 
$r_* \simeq (\ell^2/2 r_H){\ln}((r-r_H)/r_H)$.
The first and second terms of the right-hand side of Eq.~\eqref{generalsols} describe an ingoing and outgoing wave, respectively. We impose the ingoing-wave condition at the black hole horizon:~$B=0$. We then obtain the solution that satisfies the ingoing-wave condition at the black hole horizon:
\begin{equation}
\label{ingoingsolution}
\phi(z)=Az^{-i\frac{\omega\ell^2}{2r_H}}\left(1-z\right)^{\frac{1+\sqrt{1+\mu^2\ell^2}}{2}}\!_2F_1\left(a,b;c;z\right).
\end{equation}

\subsection{Asymptotic behavior at infinity}
We next consider the asymptotic behavior at infinity. For~$\mu^2>\mu_{{\rm BF}}^2=-1/\ell^2$, we have
\begin{equation}
\begin{split}
\label{asymptoticbehavior}
\phi(r)=A_{\rm I}\left(\frac{r_H}{r}\right)^{1-\sqrt{1+\mu^2\ell^2}}\left[1+\mathcal{O}(1/r^2)\right]
+ A_{\rm II}\left(\frac{r_H}{r}\right)^{1+\sqrt{1+\mu^2\ell^2}}\Xi(r),
\end{split}
\end{equation}
where the function~$\Xi$ behaves as\footnote{As will be seen later, the function~$\Xi$ has no logarithmic terms when imposing the Dirichlet boundary condition at infinity. }
\begin{equation}
\label{functionXi}
\Xi(r)=\begin{cases}&\ln r+\mathcal{O}(1),~~~-a-b+c={\rm negative~integer},\\
&1+\mathcal{O}(1/r^2),~~-a-b+c\neq{\rm negative~integer},
\end{cases}
\end{equation}
and the constants~$A_{\rm I}$, $A_{\rm II}$ are given by
\begin{equation}
\label{Am}
A_{\rm I}=A\frac{\Gamma(c)\Gamma(a+b-c)}{\Gamma(a)\Gamma(b)},
\end{equation}
and
\begin{equation}
\label{Ap}
A_{\rm II}=\begin{cases}&A\dfrac{(-1)^{1+a+b-c}\Gamma(c)}{\Gamma(-a+c)\Gamma(-b+c)},~~-a-b+c={\rm negative~integer},\\
&A\dfrac{\Gamma(c)\Gamma(-a-b+c)}{\Gamma(-a+c)\Gamma(-b+c)},~~-a-b+c\neq{\rm negative~integer}.
\end{cases}
\end{equation}
For~$\mu^2=\mu^2_{\rm BF}$, the asymptotic behavior is
\begin{equation}
\begin{split}
\label{asymptoticbehaviorBF}
\phi(r)=&A_{\rm I, BF}\frac{r_H}{r}+A_{\rm II, BF}\frac{r_H}{r}\ln\left(\frac{r_H}{r}\right)+\mathcal{O}\left(1/r^3\right),
\end{split}
\end{equation}
where
\begin{equation}
\label{tApAm}
A_{\rm I, BF}=-A\frac{2\gamma+\psi(a)+\psi(b)}{\Gamma(a)\Gamma(b)}\Gamma(a+b),~~A_{\rm II, BF}=-2A\frac{\Gamma(a+b)}{\Gamma(a)\Gamma(b)}.
\end{equation}
Here, we have used $\psi(\xi)\equiv\frac{d}{d\xi}\ln\Gamma(\xi)$ for~$\xi\in\mathbb{C}$ and the Euler number~$\gamma\equiv-\psi(1)\simeq0.5772$. In the above discussion, some formulas for the hypergeometric functions, which are given in Appendix~\ref{appendix:hypergeometricF}, are used.

\subsection{QNM boundary condition at infinity and QNM frequencies}
We consider four cases: the Dirichlet, Neumann, Robin, Dirichlet-Neumann boundary conditions. We follow the definition of the boundary conditions at infinity on~\cite{Ishibashi:2004wx}. In the following, we summarize each boundary condition and the corresponding QNM frequencies.

\subsubsection{Dirichlet boundary condition}
The Dirichlet boundary condition requires that $A_{\rm I}$ in Eq.~\eqref{Am} vanishes. Because the Gamma function has no zeros,  this condition corresponds to  $a=-n$ or~$b=-n$, where $a$, $b$ are given in Eq.~\eqref{abc} and $n$ is a nonnegative integer, then $A_{\rm I}\propto1/\Gamma(-n)=0$.  The imposition of the Dirichlet boundary condition determines the QNM frequencies,
\begin{equation}
\label{DBC}
\omega_{\rm D} =\pm\frac{m}{\ell}-i\frac{r_H}{\ell^2}\left(2n+1+\sqrt{1+\mu^2\ell^2}\right), ~~\text{(Dirichlet~B.C.)}
\end{equation}
Physically, the nonnegative integer~$n$ represents an index of overtones. 

Note that the expression of the QNM has no logarithmic term,
\begin{equation}
\phi=A\left(1-\frac{r_H^2}{r^2}\right)^{-i \frac{\omega_{\rm D}\ell^2}{2r_H}}\left(\frac{r_H}{r}\right)^{1+\sqrt{1+\mu^2\ell^2}}\sum_{k=0}^n\frac{\left(a\right)_k\left(b\right)_k}{k!\left(c\right)_k}\left(1-\frac{r_H^2}{r^2}\right)^k,
\end{equation}
where $(\xi)_k\equiv \Gamma(\xi+k)/\Gamma(\xi)$ for $\xi\in \mathbb{C}$.
This is because the logarithmic terms as in Eq.~\eqref{asymptoticbehavior} for the case, where $-a-b+c$ is a negative integer, cancel out by the imposition of the Dirichlet boundary condition~$A_{\rm I}=0$, i.e., $a=-n$ or $b=-n$, as can be seen from the formula for the hypergeometric function in Appendix~\ref{appendix:hypergeometricF}.

\subsubsection{Neumann boundary condition}
The Neumann boundary condition requires that $A_{\rm II}$ in Eq.~\eqref{Ap} vanishes. That corresponds to~$c-a=-n$ or~$c-b=-n$, where $a$, $b$, $c$ are given in Eq.~\eqref{abc} and $n$ is a nonnegative integer. Then, the QNM frequencies are determined,
\begin{equation}
\label{NBC}
\omega_{\rm N}=\pm\frac{m}{\ell}-i\frac{r_H}{\ell^2}\left(2n+1-\sqrt{1+\mu^2\ell^2}\right).
 ~~\text{ (Neumann~B.C.)}
\end{equation}
The nonnegative integer~$n$ corresponds to an index of overtones. As can be seen from this expression, for $\mu^2\ell^2> 0$, the imaginary part of $\omega_{\rm N}$ can be positive, indicating the existence of unstable modes due to the presence of a non-normalizable mode.\footnote{We call~$\Phi$ a normalizable mode if and only if the norm~$|\Phi|^2\equiv-i\int_{r_H}^{\infty} dx^3\sqrt{-g}g^{tt}\left(\Phi^*\partial_t\Phi-\Phi\partial_t\Phi^*\right)$ is finite, where~$^*$ denotes the complex conjugate. If not, we call~$\Phi$ a non-normalizable mode. In the context of AdS/CFT correspondence, a non-normalizable mode plays a role in ``source" of dual theories.}
The QNM takes the form, 
\begin{equation}
\phi=A\left(1-\frac{r_H^2}{r^2}\right)^{-i \frac{\omega_{\rm N}\ell^2}{2r_H}}\left(\frac{r_H}{r}\right)^{1-\sqrt{1+\mu^2\ell^2}}\sum_{k=0}^n\frac{\left(c-a\right)_k\left(c-b\right)_k}{k!\left(c\right)_k}\left(1-\frac{r_H^2}{r^2}\right)^k.
\end{equation}

\subsubsection{Robin boundary condition}

In the case of $\mu^2\ell^2>-1$, the Robin boundary condition corresponds to~$A_{\rm II}/ A_{\rm I}=\kappa$~($\kappa\in\mathbb{R}$), which includes the Dirichlet boundary condition~$\kappa=\infty$, i.e., $A_{\rm I}=0$, and the Neumann boundary condition~$\kappa=0$, i.e., $A_{\rm II}=0$. It is difficult to obtain the analytic expressions for the QNM frequencies except for the Dirichlet and Neumann boundary conditions. From the condition~$A_{\rm II}/ A_{\rm I}=\kappa$, the QNM frequency for a fixed $\kappa$ can be obtained numerically (see  Appendix~\ref{appendix:QNMfrequencies}).

For $\mu^2\ell^2=-1$, the Robin boundary condition corresponds to $A_{\rm II, BF}/A_{\rm I, BF}=1/\kappa_{{\rm BF}}$~($\kappa_{{\rm BF}}\in\mathbb{R}$). We numerically calculate the QNM frequency for a fixed~$\kappa$ from the condition~$A_{\rm II, BF}/A_{\rm I, BF}=1/\kappa_{{\rm BF}}$ (see Appendix~\ref{appendix:QNMfrequencies}).
We note that the QNM frequency for the~$\kappa_{\rm{BF}}=-\infty$ case can be analytically derived as shown below.

\subsubsection{Dirichlet-Neumann boundary condition}
In the case of $\mu^2\ell^2=-1$,  the Dirichlet-Neumann boundary condition is the simultaneous imposition of the Dirichlet and Neumann boundary condition~\cite{Ishibashi:2004wx}, which corresponds to $\kappa_{\rm BF}=-\infty$ in $A_{\rm II, BF}/A_{\rm I, BF}=1/\kappa_{{\rm BF}}$. The QNM frequency is calculated to
\begin{equation}
\label{DNBCBF}
\omega_{\rm DN}=\pm\frac{m}{\ell}-i\frac{r_H}{\ell^2}\left(2n+1\right).
 ~~\text{ (Dirichlet-Neumann~B.C.)}
\end{equation}
We note that this corresponds to the case of Eqs.~\eqref{DBC} and~\eqref{NBC} with $\mu^2\ell^2=-1$. The QNM takes the form, 
\begin{equation}
\label{QNMexpressionwithDN}
\phi=A\left(1-\frac{r_H^2}{r^2}\right)^{-i \frac{\omega_{\rm DN}\ell^2}{2r_H}}\left(\frac{r_H}{r}\right)\sum_{k=0}^n\frac{\left(c-a\right)_k\left(c-b\right)_k}{k!\left(c\right)_k}\left(1-\frac{r_H^2}{r^2}\right)^k.
\end{equation}

\section{Mass ladder operators in the BTZ spacetime}
\label{Sec:MLopandBCs}
In this section, we introduce ladder operators associated with the spacetime conformal symmetry of the BTZ spacetime~\eqref{BTZ} according to~\cite{Cardoso:2017qmj,Cardoso:2017egd,Katagiri:2021scx,Katagiri:2021xig}. 
In the BTZ spacetime, 
there are four independent closed conformal Killing vectors,
\begin{equation}
\begin{split}
\label{ckv}
\zeta_{0}=&e^{\frac{r_H}{\ell^2}t}\left(\frac{1}{\sqrt{r^2-{r_H}^2}}\partial_t-\frac{r\sqrt{r^2-{r_H}^2}}{\ell^2 r_H}\partial_r\right),\\
\zeta_{1}=&e^{-\frac{r_H}{\ell^2}t}\left(\frac{1}{\sqrt{r^2-{r_H}^2}}\partial_t+\frac{r\sqrt{r^2-{r_H}^2}}{\ell^2 r_H}\partial_r\right),\\
\zeta_{2}=&e^{\frac{r_H}{\ell}\varphi}\left(\frac{r^2-{r_H}^2}{\ell r_H}\partial_r+\frac{1}{r}\partial_\varphi\right),\\
\zeta_{3}=&e^{-\frac{r_H}{\ell}\varphi}\left(-\frac{r^2-{r_H}^2}{\ell r_H}\partial_r+\frac{1}{r}\partial_\varphi\right).
\end{split}
\end{equation}
Those vectors satisfy the conformal Killing equations $\nabla_{(\mu}\zeta_{\nu)} = 
(1/3)(\nabla_\alpha \zeta^\alpha)
 g_{\mu \nu}$ and 
the closed condition $\nabla_{[\mu}\zeta_{\nu]} = 0$.
We define \textit{mass ladder operators} in the static BTZ spacetime,
\begin{equation}
D_{i,k}:=\mathcal{L}_{\zeta_i}-\frac{k}{3}\nabla_\mu\zeta_{i}^\mu,
\label{massladderdef}
\end{equation}
where
$i$ runs $0, 1, 2, 3$, $k\in\mathbb{R}$ is a parameter which is related to the mass squared of the Klein-Gordon field, and 
$\mathcal{L}_{\zeta_i}$ is the Lie derivative along the closed conformal Killing vectors~\eqref{ckv}. 
The explicit forms of the mass ladder operators \eqref{massladderdef} become
\begin{equation}
\begin{split}
\label{massladder}
D_{0,k}=&e^{\frac{r_H}{\ell^2}t}\left(\frac{1}{\sqrt{r^2-{r_H}^2}}\partial_t-\frac{r\sqrt{r^2-{r_H}^2}}{\ell^2 r_H}\partial_r+k\frac{\sqrt{r^2-{r_H}^2}}{\ell^2r_H}\right),\\
D_{1,k}=&e^{-\frac{r_H}{\ell^2}t}\left(\frac{1}{\sqrt{r^2-{r_H}^2}}\partial_t+\frac{r\sqrt{r^2-{r_H}^2}}{\ell^2 r_H}\partial_r-k\frac{\sqrt{r^2-{r_H}^2}}{\ell^2r_H}\right),\\
D_{2,k}=&e^{\frac{r_H}{\ell}\varphi}\left(\frac{r^2-{r_H}^2}{\ell r_H}\partial_r+ \frac{1}{r}\partial_\varphi-k\frac{r}{\ell r_H}\right),\\
D_{3,k}=&e^{-\frac{r_H}{\ell}\varphi}\left(-\frac{r^2-{r_H}^2}{\ell r_H}\partial_r+\frac{1}{r}\partial_\varphi+k\frac{r}{\ell r_H}\right).
\end{split}
\end{equation}
For these mass ladder operators, commutation relations which act on scalar fields,
\begin{equation}
\label{commutationrelation}
\left[\nabla_\mu\nabla^\mu, D_{i,k}\right] = -\frac{2k+1}{\ell^2}D_{i,k} +\frac{2}{3}(\nabla_\mu \zeta_i^\mu)\left[\nabla_\mu\nabla^\mu -\frac{k (k+2)}{\ell^2}\right],
\end{equation}
hold, where~$\nabla_\mu\nabla^\mu$ is the d'Alembertian in the static BTZ spacetime.

Acting the commutation relation~\eqref{commutationrelation} on some smooth function~$\Phi(t,r,\varphi)$, we obtain
a relation,
\begin{equation}
\begin{split}
\label{massladdern}
&D_{i,k-2}\left[\nabla_\mu\nabla^\mu-\frac{k(k+2)}{\ell^2}\right]\Phi=\left[\nabla_\mu\nabla^\mu -\frac{(k-1)(k+1)}{\ell^2}\right]D_{i,k}\Phi.
\end{split}
\end{equation}
If $\Phi$ is a solution of the massive Klein-Gordon equation~\eqref{EOM} with mass squared~$\mu^2 \ell^2= k(k+2)$, i.e., $\Phi$ satisfies
\begin{equation}
\left[\nabla_\mu\nabla^\mu -\frac{k(k+2)}{\ell^2}\right]
\Phi =0,
\end{equation}
the left-hand side of Eq.~\eqref{massladdern} vanishes, thereby yielding
\begin{equation}
\begin{split}
\label{massshift}
\left[\nabla_\mu\nabla^\mu -\frac{(k-1)(k+1)}{\ell^2}\right]
D_{i,k}\Phi=0. 
\end{split}
\end{equation}
This shows that the mass ladder operators map a solution of the massive Klein-Gordon equation with mass squared~$\mu^2\ell^2=k(k+2)$ into that with different mass squared~$\mu^2\ell^2=(k-1)(k+1)$.

The mass squared with $\mu^2 \ge \mu_{\rm BF}^2$
can be parametrized as
\begin{equation}
\mu^2\ell^2=\nu(\nu+2), ~
\end{equation}
where $\nu \in\mathbb{R}$.\footnote{
We note that the parameter $\nu$ with $\nu \ge -1$ can represent all mass squared with $\mu^2 \ge \mu^2_{\rm BF}$. For $\nu \ge -1$, $\nu$ can be written by $\nu = -1 + \sqrt{1 + \mu^2\ell^2}$.
In this paper, we mainly focus on $\nu \ge -1$.}
For a given parameter~$\nu$, there are two solutions of $\nu(\nu+2) = k(k+2)$ with respect to $k$,
which we denote by $k_\pm$,
\begin{equation}
\label{kpm}
k_{+}=-2-\nu,~~k_{-}=\nu.
\end{equation}
\begin{figure}[htbp]
\centering
\includegraphics[scale=0.8] {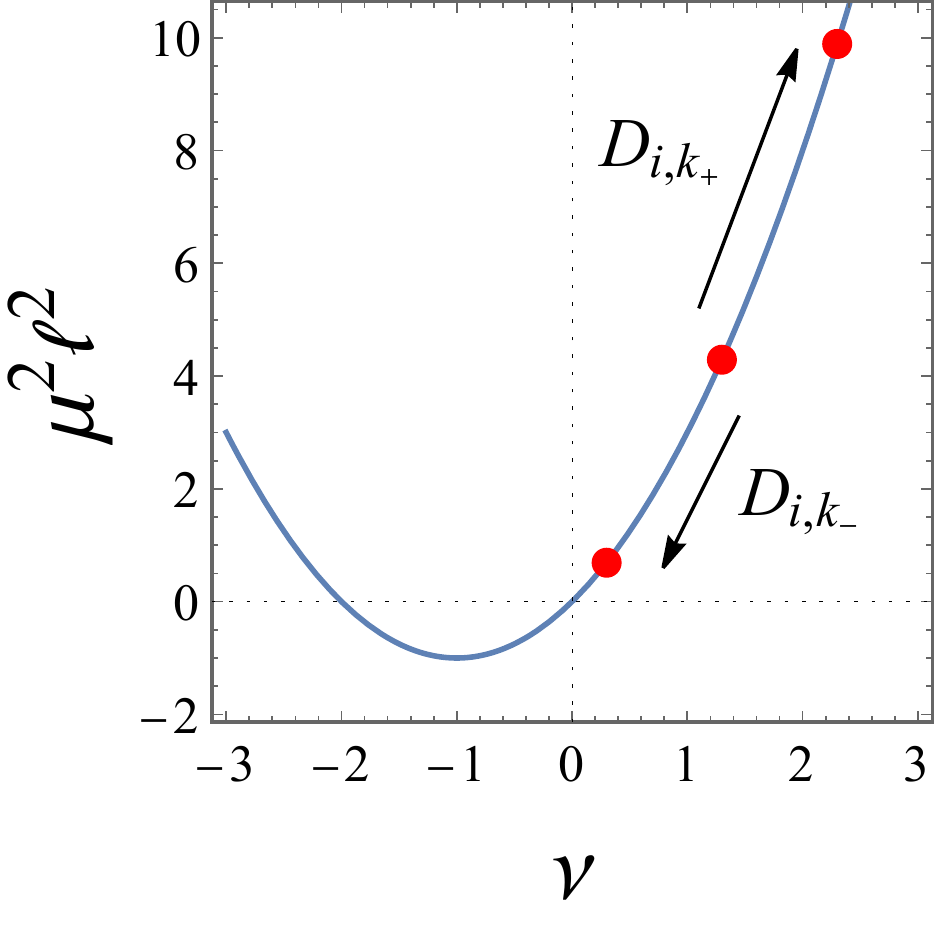}
\caption{Mass ladder operators~$D_{i,k_{\pm}}$ shift $\nu$ in the mass squared~$\mu^2\ell^2=\nu(\nu+2)$ into $\nu\to \nu\pm1$. If the original parameter is $\nu=1.3$, the parameter is shifted into $\nu=0.3$ by $D_{i,k_-}$ and into $\nu=2.3$ by $D_{i,k_+}$.\label{Figure:MassShift}} \end{figure}
Thus, there exist two different mass ladder operators,~$D_{i,k_+}$, $D_{i,k_-}$.
The operators~$D_{i,k_+}$ shift the mass squared~$\mu^2\ell^2=\nu(\nu+2)$ of the massive Klein-Gordon field to~$(\nu+1)(\nu +3)$, while the operators~$D_{i,k_-}$ shift it to~$(\nu-1)(\nu+1)$,
i.e., $D_{i,k_+}$ and $D_{i,k_-}$ shift the parameter $\nu$ to $\nu +1$ and $\nu-1$, respectively.
Depending on the value of $\mu^2$ or $\nu$, 
the mass ladder operators make mass squared raise or lower. Figure~\ref{Figure:MassShift} shows the shift of $\nu$ by the mass ladder operators~$D_{i,k_\pm}$.\footnote{Acting $D_{i,k_-}$ on the field with $-1\le \nu<-1/2$ makes the mass squared raise. In the case of $\nu=-1/2$, that keeps the mass squared.}

The above discussion shows that the mass ladder operators $D_{i,k_\pm}$ can shift the parameter~$\nu$ in the mass squared $\mu^2\ell^2=\nu(\nu+2)$ to $\nu \pm 1$.
This implies that acting the mass ladder operators several times,
the parameter $\nu$ can be shifted to $\nu \pm N$ with a positive integer~$N$.
For example, the operator
$D_{i_{N},\nu-N+1}\cdots D_{i_2,\nu-1} D_{i_1,\nu}$ can shift the mass squared from
$\mu^2\ell^2=\nu(\nu+2)$ to $(\nu-N)(\nu+2-N)$.
Another interesting example is that 
the operator $D_{i_2,-1-\nu} D_{i_1,\nu}$ becomes 
a symmetry operator of the Klein-Gordon equation 
with the mass squared
$\mu^2\ell^2=\nu(\nu+2)$, i.e., $D_{i_2,-1-\nu} D_{i_1,\nu}\Phi$ becomes the solution of
the Klein-Gordon equation with the same mass squared $\mu^2\ell^2=\nu(\nu+2)$.

Because $\zeta_0$ and $\zeta_1$ in Eq.~\eqref{ckv} 
are regular vector fields, 
the mass ladder operators $D_{0, k_\pm}$ and $D_{1, k_\pm}$ map a regular solution of the Klein-Gordon equation 
into a regular solution of that with different mass squared.
On the other hand,
$\zeta_2$ and $\zeta_3$ in Eq.~\eqref{ckv} 
are not globally regular vector fields in the BTZ spacetime due to the factor $e^{\pm r_H\varphi/\ell}$,
and the corresponding mass ladder operators $D_{2, k_\pm}$ and $D_{3, k_\pm}$ are also not globally regular.
Thus, we mainly focus on $D_{0, k_\pm}$ and $D_{1, k_\pm}$ in the following sections.\footnote{In fact, the multiple actions of the mass ladder operators $D_{2, k_\pm}$ and $D_{3, k_\pm}$
can be regular, and this case will be briefly discussed in Sec.~\ref{Sec:multipleactionsofD2D3}.}

\section{Mass ladder operators and QNM boundary conditions}
\label{Sec:MassladderandQNMBC}
In this section, we investigate the relation among the mass ladder operators~$D_{0,k}$, $D_{1,k}$ and 
asymptotic behaviors of the Klein-Gordon fields in the cases of the Dirichlet, Neumann, and Dirichlet-Neumann boundary conditions. We discuss other mass ladder operator cases, $D_{2,k}$ and $D_{3,k}$, in Appendix~\ref{Appendix:D2D3cases}; the Robin boundary condition cases in Appendix~\ref{Appendix:otherBCcase}. If the mass ladder operators keep the QNM boundary conditions, they map a QNM into another QNM appropriately. We  parametrize the mass squared of the Klein-Gordon field as $\mu^2\ell^2 = \nu(\nu + 2)$ for a real parameter~$\nu\in\mathbb{R}$. In this section, we assume $\nu\ge-1$, then $\nu=-1+\sqrt{1+\mu^2\ell^2}$.

The QNMs obtained in Sec.~\ref{Sec:QNM} are written in the form,
\begin{equation}
\label{PhiQNM}
\Phi = A 
\left(1-\frac{r_H^2}{r^2}\right)^{-i\frac{\ell^2}{2r_H}\omega}\left(\frac{r_H}{r}\right)^{2+\nu}\!_2F_1\left(a,b;c;1-{r_H^2}/{r^2}\right) e^{-i\omega t+im\varphi},
\end{equation}
where $\omega$ is the QNM frequency, and $a$,~$b$,~$c$ are given in Eq.~\eqref{abc}. 
The asymptotic behavior near the horizon is
\begin{equation}
\label{asymptoticbehaviornearBH}
\Phi|_{r\simeq r_H}=2^{-i\frac{\ell^2}{2r_H}\omega }A \left(\frac{r-r_H}{r_H}\right)^{-i\frac{\ell^2}{2r_H}\omega }\left[1+\mathcal{O}(r-r_H)\right]e^{-i\omega  t+im\varphi}.
\end{equation}
Note $r-r_H\simeq r_H e^{2r_Hr_*/\ell^2}$ in terms of the tortoise coordinate~$r_*$. For $\nu>-1$, the asymptotic behaviors at infinity are
\begin{align}
\label{asymptoticbehaviornearD}
\Phi|_{r\simeq \infty}&= A_{\rm II}\left(\frac{r_H}{r}\right)^{2+\nu}
\left[1+\mathcal{O}(1/r^2)\right]e^{-i\omega  t+im\varphi},~~{\rm (Dirichlet~B.C.)}\\
\Phi|_{r\simeq \infty}&=
A_{\rm I}\left(\frac{r_H}{r}\right)^{-\nu}\left[1+\mathcal{O}(1/r^2)\right]e^{-i\omega  t+im\varphi}.~~{\rm (Neumann~B.C.)}
\label{asymptoticbehaviornearN}\end{align}
Here, the coefficients~$A_{\rm II}$ and~$A_{\rm I}$ are, respectively, given in Eqs.~\eqref{Ap} and~\eqref{Am}. For $\nu=-1$, the asymptotic behavior with the Dirichlet-Neumann boundary condition at infinity is
\begin{equation}
\Phi|_{r\simeq \infty}^{(\nu =-1)}= 
A_{\rm II, BF}\left[\frac{r_H}{r}+\mathcal{O}(1/r^3)\right]e^{-i\omega  t+im\varphi},~~{\text {\rm (Dirichlet-Neumann~B.C.)}}
\label{asymptoticbehaviornearDN}
\end{equation}
where $A_{\rm II, BF}$ is given in Eq.~\eqref{tApAm}. The power of the leading behavior corresponds to the limit of $\nu\to-1$ of that in Eqs.~\eqref{asymptoticbehaviornearD} or~\eqref{asymptoticbehaviornearN}.

\subsection{At the horizon}
We consider the asymptotic behavior of a mapped solution near the horizon. Acting the mass ladder operators~$D_{0,k_\pm}$ and~$D_{1,k_\pm}$ on the exact solution~\eqref{PhiQNM}, the asymptotic behaviors at~$r\to r_H$ are calculated to
\begin{equation}
\begin{split}
\label{DPsi}
D_{0,k_\pm}\Phi =& c_{0,k_\pm}\left(\frac{r-r_H}{r_H}\right)^{-i\frac{\ell^2}{2r_H}\left(\omega +i\frac{r_H}{\ell^2}\right)}\left[1+\mathcal{O}(r-r_H)\right]e^{-i\left(\omega +i\frac{r_H}{\ell^2}\right)t+im\varphi},\\
D_{1,k_\pm}\Phi  =& c_{1,k_\pm} \left(\frac{r-r_H}{r_H}\right)^{-i\frac{\ell^2}{2r_H}\left(\omega -i\frac{r_H}{\ell^2}\right)}\left[1+\mathcal{O}(r-r_H)\right]e^{-i\left(\omega -i\frac{r_H}{\ell^2}\right)t+im\varphi}.
\end{split}
\end{equation}
The explicit forms of the coefficients~$c_{0,k_\pm}$ and~$c_{1,k_\pm}$ are given in Appendix~\ref{appendix:Coefficients}. Comparing these results with Eq.~\eqref{asymptoticbehaviornearBH}, it can be seen that $D_{0,k_\pm}\Phi $ are ingoing waves with the frequency~$\tilde{\omega} =\omega + i r_H/\ell^2$ and  $D_{1,k_\pm}\Phi $ are those with $\tilde{\omega} =\omega - i r_H/\ell^2$. Thus, the mass ladder operators keep the ingoing-wave condition.

\subsection{At infinity: Originally from the QNM with the Dirichlet boundary condition
}
\label{Sec:DirichletBC}
We consider the action of the mass ladder operators on $\Phi$, which satisfies the Dirichlet boundary condition. Acting the mass ladder operators~$D_{0,k_+}$ and~$D_{1,k_+}$ on the exact solution~\eqref{PhiQNM} with the Dirichlet boundary condition, the asymptotic behaviors at infinity are
\begin{equation}
\begin{split}
\label{DkrphiD}
D_{0,k_+}\Phi &=c_{0,k_+}^{\rm (D)}\left(\frac{r_H}{r}\right)^{3+\nu}
\left[1+\mathcal{O}\left(1/r^2\right)\right]~e^{-i\left(\omega +i\frac{r_H}{\ell^2}\right)t+im\varphi},\\
D_{1,k_+}\Phi &=c_{1,k_+}^{\rm (D)}\left(\frac{r_H}{r}\right)^{3+\nu}\left[1+\mathcal{O}\left(1/r^2\right)\right]~e^{-i\left(\omega -i\frac{r_H}{\ell^2}\right)t+im\varphi}.
\end{split}
\end{equation}
The explicit forms of the coefficients~$c_{0,k_+}^{\rm (D)}$ and~$c_{1,k_+}^{\rm (D)}$ are given in Appendix~\ref{appendix:Coefficients}.
Comparing these results with Eq.~\eqref{asymptoticbehaviornearD}, it can be seen that these correspond to the asymptotic forms of a solution with mass squared~$\mu^2\ell^2=\tilde{\nu}(\tilde{\nu}+2)$ with $\tilde{\nu}=\nu+1$, which satisfies the Dirichlet boundary condition. Thus, the mass ladder operators~$D_{0,k_+}$ and $D_{1,k_+}$ keep the Dirichlet boundary condition. 

Acting the mass ladder operators~$D_{0,k_-}$ and~$D_{1,k_-}$ on the exact solution~\eqref{PhiQNM} with the Dirichlet boundary condition, the asymptotic behaviors at infinity are
\begin{equation}
\begin{split}
\label{DklphiD}
D_{0,k_-}\Phi &=c_{0,k_-}^{\rm (D)}\left(\frac{r_H}{r}\right)^{1+\nu}\left[1+\mathcal{O}\left(1/r^2\right)\right]~e^{-i\left(\omega +i\frac{r_H}{\ell^2}\right)t+im\varphi},\\
D_{1,k_-}\Phi &=c_{1,k_-}^{\rm (D)}
\left(\frac{r_H}{r}\right)^{1+\nu}\left[1+\mathcal{O}\left(1/r^2\right)\right]~e^{-i\left(\omega -i\frac{r_H}{\ell^2}\right)t+im\varphi}.
\end{split}
\end{equation}
The explicit forms of the coefficients~$c_{0,k_-}^{\rm (D)}$ and~$c_{1,k_-}^{\rm (D)}$ are given in Appendix~\ref{appendix:Coefficients}.
Comparing these results with Eqs.~\eqref{asymptoticbehaviornearD}-\eqref{asymptoticbehaviornearDN}, we can read off the boundary conditions at infinity depending on the value of $\nu$.  The asymptotic forms of a solution with mass squared~$\mu^2\ell^2=\tilde{\nu}(\tilde{\nu}+2)$ with $\tilde{\nu}=\nu-1$, which satisfies the Dirichlet boundary condition for $\nu>0$. Thus, for $\nu>0$, the mass ladder operators~$D_{0,k_-}$ and~$D_{1,k_-}$  also keep the Dirichlet boundary condition. For $\nu=0$, the mass ladder operators~$D_{0,k_-}$ and~$D_{1,k_-}$ change into the Dirichlet-Neumann boundary condition as can be seen by comparing with Eq.~\eqref{asymptoticbehaviornearDN}.
For $-1<\nu<0$, the asymptotic behaviors in Eq.~\eqref{DklphiD} are the asymptotic forms of a solution with mass squared~$\mu^2\ell^2=\tilde{\nu}(\tilde{\nu}+2)$ with $\tilde{\nu}=|\nu|-1$, which satisfies the Neumann boundary condition: $A_{\rm II}=0$. Note that $\tilde{\nu}$ satisfies $-1<\tilde{\nu}<0$ for $-1<\nu<0$. Thus, remarkably, for $-1<\nu<0$, the mass ladder operators~$D_{0,k_-}$ and $D_{1,k_-}$ change the Dirichlet boundary condition into the Neumann boundary condition. 

\subsection{At infinity: Originally from the QNM with the Neumann boundary condition}
\label{Sec:NeumannBC}
We act the mass ladder operators on $\Phi $, which satisfies the Neumann boundary condition. Acting the mass ladder operators~$D_{0,k_+}$ and~$D_{1,k_+}$ on the exact solution~\eqref{PhiQNM} with the Neumann boundary condition, the asymptotic behaviors at infinity are
\begin{equation}
\begin{split}
\label{DkrphiN}
D_{0,k_+}\Phi &=c_{0,k_+}^{\rm (N)}\left(\frac{r_H}{r}\right)^{-1-\nu}\left[1+\mathcal{O}(1/r^2)\right]e^{-i\left(\omega +i\frac{r_H}{\ell^2}\right)t+im\varphi},\\
D_{1,k_+}\Phi &=c_{1,k_+}^{\rm (N)}\left(\frac{r_H}{r}\right)^{-1-\nu}\left[1+\mathcal{O}(1/r^2)\right]e^{-i\left(\omega -i\frac{r_H}{\ell^2}\right)t+im\varphi}.
\end{split}
\end{equation}
The explicit forms of the coefficients~$c_{0,k_+}^{\rm (N)}$ and~$c_{1,k_+}^{\rm (N)}$ are given in Appendix~\ref{appendix:Coefficients}. These are the asymptotic forms of a solution with mass squared~$\mu^2\ell^2=\tilde{\nu}(\tilde{\nu}+2)$ with $\tilde{\nu}=\nu+1$, which satisfies the Neumann boundary condition. Thus, the mass ladder operators~$D_{0,k_+}$ and $D_{1,k_+}$ keep the Neumann boundary condition.

Acting the mass ladder operators~$D_{0,k_-}$ and~$D_{1,k_-}$ on the exact solution~\eqref{PhiQNM} with the Neumann boundary condition, the asymptotic behaviors at infinity are
\begin{equation}
\begin{split}
\label{DklphiN}
D_{0,k_-}\Phi &=c_{0,k_-}^{\rm (N)}\left(\frac{r_H}{r}\right)^{1-\nu}\left[1+\mathcal{O}(1/r^2)\right]e^{-i\left(\omega +i\frac{r_H}{\ell^2}\right)t+im\varphi},\\
D_{1,k_-}\Phi &=c_{1,k_-}^{\rm (N)}\left(\frac{r_H}{r}\right)^{1-\nu}\left[1+\mathcal{O}(1/r^2)\right]e^{-i\left(\omega -i\frac{r_H}{\ell^2}\right)t+im\varphi}.
\end{split}
\end{equation}
The explicit forms of the coefficients~$c_{0,k_-}^{\rm (N)}$ and~$c_{1,k_-}^{\rm (N)}$ are given in Appendix~\ref{appendix:Coefficients}.  Comparing these results with Eqs.~\eqref{asymptoticbehaviornearD}-\eqref{asymptoticbehaviornearDN}, we can read off the boundary conditions at infinity depending on the value of $\nu$. For $\nu>0$, the asymptotic forms of a solution with mass squared~$\mu^2\ell^2=\tilde{\nu}(\tilde{\nu}+2)$ with $\tilde{\nu}=\nu-1$, which satisfies the Neumann boundary condition. Thus, for $\nu>0$, the mass ladder operators~$D_{0,k_-}$ and~$D_{1,k_-}$  also keep the Neumann boundary condition. For $\nu=0$, the mass ladder operators~$D_{0,k_-}$ and~$D_{1,k_-}$ change into the Dirichlet-Neumann boundary condition as can be seen by comparing with Eq.~\eqref{asymptoticbehaviornearDN}.
For $-1<\nu<0$, the asymptotic behaviors in Eq.~\eqref{DklphiN} are the asymptotic forms of a solution with mass squared~$\mu^2\ell^2=\tilde{\nu}(\tilde{\nu}+2)$ with $\tilde{\nu}=|\nu|-1$, which satisfies the Dirichlet boundary condition. Note that $\tilde{\nu}$ satisfies $-1<\tilde{\nu}<0$ for $-1<\nu<0$. It should also be noted that, for $-1<\nu<0$, the mass ladder operators~$D_{0,k_-}$ and $D_{1,k_-}$ change the Neumann boundary condition into the Dirichlet boundary condition.

\subsection{At infinity: Originally from the QNM with the Dirichlet-Neumann boundary condition}
We act the mass ladder operators~$D_{0,-1}$ and~$D_{1,-1}$ on~$\Phi$, which satisfies the Dirichlet-Neumann boundary condition.\footnote{Because of $k_+=k_-=-1$ in Eq.~\eqref{kpm}, both the mass ladder operators~$D_{k_+}$ and $D_{k_-}$ make the mass squared raise as stated at the end of Sec.~\ref{Sec:MLopandBCs}.} Acting them on the exact solution~\eqref{PhiQNM} with the Dirichlet-Neumann boundary condition, the asymptotic behaviors at infinity are
\begin{equation}
\begin{split}
\label{DkphiDN}
D_{0,-1}\Phi &=c_{0,-1}^{\rm (DN)}\left(\frac{r_H}{r}\right)^{2}\left[1+\mathcal{O}(1/r^2)\right]e^{-i\left(\omega +i\frac{r_H}{\ell^2}\right)t+im\varphi},\\
D_{1,-1}\Phi &=c_{1,-1}^{\rm (DN)}\left(\frac{r_H}{r}\right)^{2}\left[1+\mathcal{O}(1/r^2)\right]e^{-i\left(\omega -i\frac{r_H}{\ell^2}\right)t+im\varphi}.
\end{split}
\end{equation}
The explicit forms of the coefficients~$c_{0,-1}^{\rm (DN)}$ and~$c_{1,-1}^{\rm (DN)}$ are given in Appendix~\ref{appendix:Coefficients}. 
Comparing with Eq.~\eqref{asymptoticbehaviornearD},
these correspond to the asymptotic forms of a massless solution, which satisfies the Dirichlet boundary condition. Thus, the mass ladder operators~$D_{0,-1}$ and~$D_{1,-1}$ change the Dirichlet-Neumann boundary condition into the Dirichlet boundary condition.

\section{Changes of QNM frequencies and Interpretation}
\label{Sec:MLopandQNMs}
In the previous section, for the cases of the Dirichlet, Neumann, and Dirichlet-Neumann boundary conditions, we have seen that the single action of $D_{0,k_\pm}$ and $D_{1,k_\pm}$ maps the QNM into another QNM with one of those boundary conditions depending on the value of $\nu$~(see Tables~\ref{table:forDirichletcase},~\ref{table:forNeumanncase}, and~\ref{table:forDNcase}). In this section, we discuss physical interpretations in terms of the QNM frequency spectra. In particular, for those boundary conditions, we will see that all overtones can be generated from the fundamental modes.

The QNM frequencies with the Dirichlet, Neumann, and Dirichlet-Neumann boundary conditions are, respectively, given in Eqs.~\eqref{DBC}, ~\eqref{NBC}, and~\eqref{DNBCBF}, which are rewritten in terms of~$\nu$ as
\begin{align}
\label{DBCnu}
\omega_{\rm D}\left(\nu,n\right) =&\pm\frac{m}{\ell}-i\frac{r_H}{\ell^2}\left(2n+2+\nu\right),~~({\rm Dirichlet~B.C.)}\\
\label{NBCnu}
\omega_{\rm N} \left(\nu,n\right) =&\pm\frac{m}{\ell}-i\frac{r_H}{\ell^2}\left(2n-\nu\right),~~({\rm Neumann~B.C.)}\\
\label{DNBCnu}
\omega_{\rm DN} \left(n\right) =&\pm\frac{m}{\ell}-i\frac{r_H}{\ell^2}\left(2n+1\right),~~({\rm Dirichlet{\text-}Neumann~B.C.)}
\end{align}
where $m\in \mathbb{Z}$ is the azimuthal number and $n=0,1,2,\cdots$ is an index of overtones.
Note that the parameter~$\nu$ is related to the mass squared~$\mu^2\ell^2(\ge \mu_{\rm BF}^2\ell^2)$ as $\nu=-1+\sqrt{1+\mu^2\ell^2}$ under the assumption~$\nu\ge-1$.

\subsection{From the QNM with the Dirichlet boundary condition}
Equations~\eqref{DkrphiD} and~\eqref{DklphiD} imply that acting the mass ladder operators~$D_{0,k_\pm}$ on the QNM, the QNM frequency is shifted as $\omega\to\tilde{\omega}=\omega+ir_H/\ell^2$, i.e.,
\begin{equation}
\label{resultingDBCnu0}
\tilde{\omega} =\pm\frac{m}{\ell}-i\frac{r_H}{\ell^2}\left(2n+1+\nu\right).
\end{equation}
It also follows from Eqs.~\eqref{DkrphiD} and~\eqref{DklphiD} that acting the mass ladder operators~$D_{1,k_\pm}$, we observe~$\omega\to\tilde{\omega}=\omega-ir_H/\ell^2$, i.e.,
\begin{equation}
\label{resultingDBCnu1}
\tilde{\omega} =\pm\frac{m}{\ell}-i\frac{r_H}{\ell^2}\left(2n+3+\nu\right).
\end{equation}
These frequency shifts can be interpreted as the changes of $(\nu, n)$ by comparing with Eqs.~\eqref{DBCnu}-\eqref{DNBCnu}. We summarize the results in Table~\ref{table:forDirichletcase}. 
We should note that 
$D_{0,k_-}$ and $D_{1,k_-}$ change the boundary condition of the QNMs for $-1< \nu \le 0$.
When acting $D_{0,k_+}$ on the fundamental mode of $\Phi$, i.e., $n=0$, a trivial solution is generated: $D_{0,k_+}\Phi=0$. Namely, no ``negative overtones" are generated. This can also be understood from the fact that the coefficients~$c_{0,k_+}$ and $c_{0,k_+}^{\rm (D)}$ in Eqs.~\eqref{c01kpkm} and~\eqref{cDkp} vanish for the fundamental QNM frequency~$\omega_{\rm D}(\nu,0)$.

\begin {table}
     \caption {Change of the QNM frequencies by the mass ladder operators. The original QNM frequency is~$\omega_{\rm D}(\nu,n)$ in Eq.~\eqref{DBCnu}. For $D_{0,k_+}$, no ``negative overtones" are generated from the fundamental mode~$n=0$.\label{table:forDirichletcase}}
  \begin {tabular}{|c|c|c|c|c|} \hline
     Operators& $D_{0,k_+}$& $D_{0,k_-}$& $D_{1,k_+}$& $D_{1,k_-}$\\ \hline
   &  & $ \omega_{\rm D}(\nu-1,n)~(\nu>0)$& $$ & $ \omega_{\rm D}(\nu-1,n+1)~(\nu>0)$\\ 
  Frequencies & $\omega_{\rm D}(\nu+1,n-1)$ & $\omega_{\rm DN}(n)~(\nu=0)$& $ \omega_{\rm D}(\nu+1,n)$ & $ \omega_{\rm DN}(n+1)~(\nu=0)$ \\ 
       & & $\omega_{\rm N}(|\nu|-1,n)~(-1<\nu<0)$& & $\omega_{\rm N}(|\nu|-1,n+1)~(-1<\nu<0)$\\
      \hline
      \end {tabular}
\end {table}

\subsection{From the QNM with the Neumann boundary condition}
From Eqs.~\eqref{DkrphiN} and~\eqref{DklphiN}, acting the mass ladder operators~$D_{0,k_\pm}$ on the QNM, we obtain a shifted QNM frequency,
\begin{equation}
\tilde{\omega} =\pm\frac{m}{\ell}-i\frac{r_H}{\ell^2}\left(2n-\nu-1\right).
\end{equation}
Equations~\eqref{DkrphiN} and~\eqref{DklphiN} also imply that acting the mass ladder operators~$D_{1,k_\pm}$, we obtain
\begin{equation}
\label{resultingNBCnu1}
\tilde{\omega} =\pm\frac{m}{\ell}-i\frac{r_H}{\ell^2}\left(2n-\nu+1\right).
\end{equation}
These frequency shifts can be interpreted as the changes of $(\nu, n)$ by comparing with Eqs.~\eqref{DBCnu}-\eqref{DNBCnu}. We summarize the results in Table~\ref{table:forNeumanncase}. 
We should note that 
$D_{0,k_-}$ and $D_{1,k_-}$ change the boundary condition of the QNMs for $-1< \nu \le 0$.
When acting $D_{0,k_-}$ on the fundamental mode of $\Phi$, i.e., $n=0$, a trivial solution is generated: $D_{0,k_-}\Phi=0$; thus, no ``negative overtones" are generated. This can also be understood from the fact that the coefficients~$c_{0,k_-}$ and $c_{0,k_-}^{\rm (N)}$ in Eqs.~\eqref{c01kpkm} and~\eqref{cNkm} vanish for the fundamental QNM frequency~$\omega_{\rm N}(\nu,0)$.

\begin {table}
     \caption {Change of the QNM frequencies by the mass ladder operators. The original QNM frequency is~$\omega_{\rm N}(\nu,n)$ in Eq.~\eqref{NBCnu}. 
For $D_{0,k_-}$, no ``negative overtones" are generated from the fundamental mode~$n=0$.\label{table:forNeumanncase}}
  \begin {tabular}{|c|c|c|c|c|} \hline
     Operators& $D_{0,k_+}$& $D_{0,k_-}$& $D_{1,k_+}$& $D_{1,k_-}$\\ \hline
   &  & $ \omega_{\rm N}(\nu-1,n-1)~(\nu>0)$& $$ & $ \omega_{\rm N}(\nu-1,n)~(\nu>0)$\\ 
  Frequencies & $\omega_{\rm N}(\nu+1,n)$ & $\omega_{\rm DN}(n-1)~(\nu=0)$& $ \omega_{\rm N}(\nu+1,n+1)$ & $ \omega_{\rm DN}(n)~(\nu=0)$ \\ 
       & & $\omega_{\rm D}(|\nu|-1,n-1)~(-1<\nu<0)$& & $\omega_{\rm D}(|\nu|-1,n)~(-1<\nu<0)$\\
      \hline
      \end {tabular}
\end {table}

\subsection{From the QNM with the Dirichlet-Neumann boundary condition}
From Eq.~\eqref{DkphiDN}, acting the mass ladder operator~$D_{0,-1}$ on the QNM, the QNM frequency is shifted as $\omega\to\tilde{\omega}=\omega+ir_H/\ell^2$, i.e.,
\begin{equation}
\tilde{\omega} =\pm\frac{m}{\ell}-i\frac{r_H}{\ell^2}\left(2n\right).
\end{equation}
Equation~\eqref{DkphiDN} also implies that acting the mass ladder operator~$D_{1,-1}$, we obtain a shifted QNM frequency,
\begin{equation}
\label{resultingNBCnu1}
\tilde{\omega} =\pm\frac{m}{\ell}-i\frac{r_H}{\ell^2}\left(2n+2\right).
\end{equation}
These frequency shifts can be interpreted as the changes of $(\nu,n)$ by comparing with Eq.~\eqref{DBCnu}. We summarize the results in Table~\ref{table:forDNcase}. We should note that 
$D_{0,-1}$ and $D_{1,-1}$ change the boundary condition of the QNMs from the Dirichlet-Neumann boundary condition into the Dirichlet boundary condition.
When acting $D_{0,-1}$ on the fundamental mode of $\Phi$, i.e., $n=0$, a trivial solution is generated: $D_{0,-1}\Phi=0$; thus, no ``negative overtones" are generated. This can also be understood from the fact that the coefficients~$c_{0,-1}$ and $c_{0,-1}^{\rm (DN)}$ in Eqs.~\eqref{c01kpkm} and~\eqref{cDN} vanish for the fundamental QNM frequency~$\omega_{\rm DN}(0)$.

In fact, for the $\mu^2 = \mu_{\rm BF}^2$ case, 
we can define an additional mass ladder operator which acts on only  a fundamental mode as
\begin{align}
D_{0}^{\rm BF} \Phi_0 := (\nabla_\mu \zeta^\mu_0) \Phi_0 = 
e^{r_H t/\ell^2} \frac{\sqrt{r^2 -r_H^2}}{\ell^2 r_H} \Phi_0,
\end{align}
where $\Phi_0$ is the fundamental mode in Eq.~\eqref{QNMexpressionwithDN} with $n = 0$.
We can easily see that $D_{0}^{\rm BF} \Phi_0$ is 
the fundamental mode of the QNM with the Neumann boundary condition 
for the massless Klein-Gordon equation.\footnote{The reason why this operator acts as a ladder operator only for the fundamental mode is as follows. 
We parametrize the mass squared around $\mu_{\rm BF}^2$ 
as $\mu^2\ell^2 = -1 + \delta^2$ with $1 \gg \delta \ge 0$, then $k_\pm$ becomes $k_\pm = -1 \pm \delta$. The additional mass ladder operator $D_{0}^{\rm BF}$ 
can be written by 
\begin{align} 
D_{0}^{\rm BF} = -\frac{3}{2}\lim_{\delta \to 0} \frac{D_{0,k_+} - D_{0,k_-}}{\delta}.
\end{align}
For $\delta >0$, if the operator $D_{0,k_+} - D_{0,k_-}$ 
acts on the fundamental mode of the QNM with the Dirichlet boundary condition, 
the mapped solution becomes the QNM with the Neumann boundary condition because the
action of $D_{0,k_+}$ on the fundamental mode vanishes (see Table~\ref{table:forDirichletcase}).
Taking the limit of $\delta \to 0$, the mapped solution becomes the 
fundamental mode of the QNM with the Neumann boundary condition 
for the massless Klein-Gordon equation.}

\begin {table}
     \caption {Change of the QNM frequencies by the mass ladder operators. The original QNM frequency is~$\omega_{\rm DN}(n)$ in Eq.~\eqref{DNBCnu}. 
For $D_{0,-1}$, no ``negative overtones" are generated from the fundamental mode~$n=0$. The operator~$D_{0}^{\rm BF}$ maps the fundamental mode into a fundamental mode of the QNM with the Neumann boundary condition for the massless Klein-Gordon equation.\label{table:forDNcase}}
  \begin {tabular}{|c|c|c|c|} \hline
     Operators&  $D_{0,-1}$& $D_{1,-1}$ &$D_0^{\rm BF}$\\ \hline
    Frequencies & $ \omega_{\rm D}(0,n-1)$ & $ \omega_{\rm D}(0,n)$ & $\omega_{{\rm DN}}(0)\to \omega_{\rm N}(0,0)$
      \\\hline
      \end {tabular}
\end {table}

\subsection{Symmetry operators from multiple actions}
Symmetry operators can be constructed from multiple actions of $D_{0,k_\pm}$ and $D_{1,k_\pm}$. For example, the actions~$D_{0,k_+-1}D_{0,k_-}$ or $D_{1,k_+-1}D_{1,k_-}$ keep mass squared as the original one and maps the QNM into a QNM appropriately. Then, the QNM frequency is shifted as~$\omega\to\omega\pm i 2r_H/\ell^2$. In the case of the Dirichlet boundary condition, in which the QNM frequency is given in Eq.~\eqref{DBCnu},
that leads to
\begin{equation}
\label{tildeomegaD0D0}
\tilde{\omega}=\pm\frac{m}{\ell}-i\frac{r_H}{\ell^2}\left[2\left(n-1\right)+2+\nu\right],
\end{equation}
for $D_{0,k_+-1}D_{0,k_-}\Phi$ and
\begin{equation}
\label{tildeomegaD1D1}
\tilde{\omega}=\pm\frac{m}{\ell}-i\frac{r_H}{\ell^2}\left[2\left(n+1\right)+2+\nu\right],
\end{equation}
for $D_{1,k_+-1}D_{1,k_-}\Phi$. It can be seen that the index of the overtone is shifted. This is also the case of the Neumann boundary condition and the Dirichlet-Neumann boundary condition. Thus, for those boundary conditions, all overtones can be generated from the fundamental modes by acting these symmetry operators multiple times.\footnote{This is not necessarily the case of the Robin boundary condition.} We note that if a symmetry operator constructed from the multiple action of the mass ladder operators
does not change the overtone number, the operator can be written by 
the multiple action of the Lie derivative with respect to the Killing vectors.

\section{Remarks}
\label{Sec: Remark}

\subsection{Parameter shifts of QNMs with purely imaginary frequency for the Robin boundary condition}
\label{Sec:InterpretationforRobinShift}
In Appendix~\ref{Appendix:otherBCcase}, we discuss changes of QNMs with the Robin boundary condition with the mass ladder operators. The QNM frequencies are shifted as $\omega \to\omega+ i r_H/\ell^2$ and $\omega \to\omega -i r_H/\ell^2$ for $D_{0,k_\pm}$ and $D_{1,k_\pm}$, respectively.
The mass ladder operators $D_{0,k_\pm}$ and $D_{1,k_\pm}$ change not only the mass squared but also the Robin boundary condition parameter~$\kappa$. In general, the resulting Robin parameter is complex even if the original parameter is real.
We find that the shifted Robin parameter takes real values 
when the original QNM frequency is purely imaginary and the original Robin parameter is real.
As can be seen in Figures~\ref{QNF} and~\ref{QNFBF} in Appendix~\ref{appendix:QNMfrequencies}, the purely imaginary~$\omega$ indeed exist. 
When acting the mass ladder operators on the QNMs with purely imaginary~$\omega$ and the real Robin parameter, 
resulting QNMs also have purely imaginary frequencies and the real Robin parameter.

\subsection{Regular solution generated by multiple actions of $D_{2,k_\pm}$ and $D_{3,k_\pm}$}
\label{Sec:multipleactionsofD2D3}
As mentioned in Sec.~\ref{Sec:MLopandBCs}, the single action of $D_{2,k_\pm}$ and $D_{3,k_\pm}$ fails to generate a globally regular solution from the QNM due to the factor~$e^{\pm r_H\varphi/\ell}$. Here, we argue that a special combination of them yields a regular solution, i.e., a QNM.
For example, the multiple action~$D_{2,k_+-1}D_{3,k_+}$ has no singular factor and changes the parameter~$\nu$ as $\nu\to\nu+2$. The QNM is appropriately mapped into a QNM with mass squared raised but the value of $\omega$ is unchanged. The resulting QNM frequencies are interpreted as
\begin{equation}
\tilde{\omega}=\pm\frac{m}{\ell}-i\frac{r_H}{\ell^2}\left[2\left(n-1\right)+2+\left(\nu+2\right)\right],
\end{equation}
for the Dirichlet boundary condition, and 
\begin{equation}
\tilde{\omega}=\pm\frac{m}{\ell}-i\frac{r_H}{\ell^2}\left[2\left(n+1\right)-\left(\nu+2\right)\right],
\end{equation}
for the Neumann boundary condition. The indices of the overtone are also shifted. 

Finally, we make a comment on symmetry operators. If we construct regular symmetry operators from $D_{2,k_\pm}$ and $D_{3,k_\pm}$, e.g., $D_{2,k_-+1} D_{3,k_+}$, 
those do not change the QNM frequencies. In fact, those operators become the trivial symmetry operators, i.e., the Lie derivative with respect to the Killing vectors.

\section{Summary}
\label{Sec:Summary}
We have studied QNMs of the massive Klein-Gordon field in the static BTZ spacetime in terms of the mass ladder operator, which is constructed from the spacetime conformal symmetry. The mass ladder operators consist of two types of operators, which play a role in making mass squared of the Klein-Gordon field raise or lower, respectively.

We have shown that the mass ladder operators keep the ingoing-wave condition at the horizon and can change boundary conditions at infinity. In the cases of the Dirichlet and Neumann boundary condition, we have seen that the mass ladder operators map the QNM into another QNM with one of those boundary conditions depending on the value of the mass squared. It has further been shown that the mass ladder operators change not only the mass squared but also the QNM frequencies. In some cases, an index of overtones is shifted. It has been revealed that all overtones can be generated from a fundamental mode with the mass ladder operators. This reminds us of the well-known results in quantum mechanical problems, i.e.,
all excited states can be derived by acting the ladder operators on the ground state wave function in the harmonic oscillator or supersymmetric system. It is also known that there is a correspondence among QNMs and bound states of the Schr\"{o}dinger equation in quantum mechanics, e.g.,~\cite{Ferrari:1984ozr,Ferrari:1984zz,Zaslavsky:1991ug,Hatsuda:2019eoj}.
Our present results suggest the existence of a further analogy between QNMs and quantum mechanical systems.

We have also investigated the case of the Robin boundary condition, which is characterized by one real parameter called the Robin parameter. In this case, the mass ladder operators keep the Robin boundary condition but change the Robin parameter. In general, the resulting Robin parameter is complex even if the original parameter is real. We have found that the shifted Robin parameter takes real values when the original QNM frequency is purely imaginary.

It is interesting to extend the current analysis to the case of rotating BTZ black holes. Test fields in the rotating BTZ black hole spacetime have been intensively investigated in, e.g.,~\cite{Natsuume:2020snz,Noda:2019mzd,  Dias:2019ery}. 
Because the rotating BTZ spacetime is also locally the anti-de Sitter spacetime in three dimensions,
the mass ladder operators can also be defined by using the spacetime conformal symmetry.
Thus, we expect that the ladder operators can be used to understand the QNM frequencies of rotating BTZ black holes. It is also interesting to study QNMs of topological black holes in~\cite{Aros:2002te}, of which geometries are locally isometric to the anti-de Sitter spacetimes in four and higher dimensions. We leave these problems for future work.

\begin{acknowledgments}
We would like to thank Tomohiro Harada, Yasuyuki Hatsuda, Takaaki Ishii and Norihiro Tanahashi for useful comments and discussions. This work was supported by Rikkyo University Special Fund for Research (T.K.) and JSPS KAKENHI Grant Nos. JP17H06360 (T.K.), JP20H04746 (M.K.) and JP22K03626 (M.K.). T.K. is also supported by VILLUM FONDEN (grant no. 37766), by the Danish Research Foundation, and under the European
Union's H2020 ERC Advanced Grant ``Black holes: gravitational engines of discovery" grant agreement no. Gravitas-101052587.

\end{acknowledgments}

\appendix

\section{Formulas of hypergeometric function}
\label{appendix:hypergeometricF}
We give the formulas for the hypergeometric functions~\cite{Abramowitz:1972} used in Sec.~\ref{Sec:QNM}. In the derivation of Eq.~\eqref{asymptoticbehavior} for $\mu^2\ell^2>-1$, if $-a-b+c=-\sqrt{1+\mu^2\ell^2}$ is a negative integer, e.g., $\mu^2=0$, we use a property, 
\begin{equation}
\begin{split}
\!_2F_1\left(a,b;c;z\right)=&\frac{\Gamma(c)\Gamma(a+b-c)}{\Gamma(a)\Gamma(b)}\left(1-z\right)^{-a-b+c}\sum_{k=0}^{a+b-c-1}\frac{\left(-a+c\right)_k\left(-b+c\right)_k}{k!\left(a+b-c+1\right)_k}(1-z)^k\\
&-\frac{(-1)^{a+b-c}\Gamma\left(c\right)}{\Gamma\left(-a+c\right)\Gamma\left(-b+
c\right)}\sum_{k=0}^{\infty}\frac{(a)_k(b)_k}{k!(a+b-c+k)!}(1-z)^k\\
&\times\left(\ln(1-z)-\psi(k+1)-\psi(a+b-c+1+k)+\psi(a+k)+\psi(b+k)\right),
\end{split}
\end{equation}
where $(\xi)_k\equiv\Gamma(\xi+k)/\Gamma(\xi)$ for~$\xi\in\mathbb{C}$.
If $-a-b+c=-\sqrt{1+\mu^2\ell^2}$ is not a negative integer, we use a property,
\begin{equation}
\begin{split}
\!_2F_1\left(a,b;c;z\right)=&\frac{\Gamma(c)\Gamma(a+b-c)}{\Gamma(a)\Gamma(b)}\left(1-z\right)^{-a-b+c} ~\!_2F_1\left(-a+c,-b+c;-a-b+c+1;1-z\right)\\
&+\frac{\Gamma(c)\Gamma(-a-b+c)}{\Gamma(-a+c)\Gamma(-b+c)}~\!_2F_1\left(a,b;a+b-c+1;1-z\right).
\end{split}
\end{equation}
In the derivation of Eq.~\eqref{asymptoticbehaviorBF} for $\mu^2\ell^2=-1$, we use a property,
\begin{equation}
\begin{split}
\!_2F_1\left(a,b;a+b;z\right)=&\frac{\Gamma(a+b)}{\Gamma\left(a\right)\Gamma\left(b\right)}\sum_{k=0}^{\infty}\frac{(a)_k(b)_k}{(k!)^2}(1-z)^k\\
&\times\left(-\ln(1-z)+2\psi(1+k)-\psi(a+k)-\psi(b+k)\right).
\end{split}
\end{equation}

For the Dirichlet boundary condition, i.e., $a=-n$ (or $b=-n$) for a nonnegative integer~$n$, the hypergeometric function takes the form of a finite polynomial,
\begin{equation}
\!_2F_1\left(-n,b;c;z\right)=\sum_{k=0}^n\frac{\left(-n\right)_k\left(b\right)_k}{\left(c\right)_k}\frac{z^k}{k!}.
\end{equation}
Note that there is a property~$_2F_1(a,b;c;z)=\!_2F_1(b,a;c;z)$; therefore, $\!_2F_1(a,-n;c;z)$ is also a finite polynomial of $z$.

\section{QNM frequencies with the Robin boundary condition}
\label{appendix:QNMfrequencies}
In this section, we numerically calculate QNM frequencies with the Robin boundary condition
due to the difficulty of the analytical derivation of them.

\subsection{ $\mu^2\ell^2>-1$ case}
For $\mu^2\ell^2>-1$, the generic boundary condition at infinity corresponds to choosing 
the values of $A_{\rm I}$ and $A_{\rm II}$ in the asymptotic behavior in Eq.~\eqref{asymptoticbehavior}.
When we solve the linear differential equation, the only meaningful information is the ratio of $A_{\rm I}$ and $A_{\rm II}$. Thus, we choose the value of $\kappa$ such that $A_{\rm II}/A_{\rm I} = \kappa~(\kappa\in\mathbb{R})$.
This generic boundary condition is called the Robin boundary condition. 
We note that the Robin boundary condition includes the Dirichlet boundary condition~$\kappa=\infty$, i.e.,~$A_{\rm I}=0$, and the Neumann boundary condition~$\kappa=0$, i.e.,~$A_{\rm II}=0$.\footnote{The Dirichlet or Neumann boundary conditions require that the leading term of~$(r/r_H)^{1-\sqrt{1+\mu^2\ell^2}}\phi$ or~$\partial_r((r/r_H)^{1-\sqrt{1+\mu^2\ell^2}}\phi)$ vanishes near~$r\to\infty$, respectively, while the Robin boundary condition requires their linear combination vanishes~\cite{Ishibashi:2004wx}. The dominant behaviors of~$(r/r_H)^{1-\sqrt{1+\mu^2\ell^2}}\phi$ and~$\partial_r((r/r_H)^{1-\sqrt{1+\mu^2\ell^2}}\phi)$ are governed by~$A_{\rm I}$ and~$A_{\rm II}$, respectively. Hence,~$A_{\rm I}=0$ and~$A_{\rm II}=0$ correspond to the Dirichlet boundary condition and the Neumann boundary condition, respectively.} In this section, we investigate a typical mass squared case in the region of $-1<\mu^2\ell^2<0$, where there are two normalizable modes. An analysis of QNMs in this case is well motivated from various contexts, e.g.,~\cite{Witten:2001ua,Hertog:2004ns,Dappiaggi:2017pbe,Katagiri:2020mvm}.

\begin{figure}[htbp]
\centering
\subfigure[QNM frequencies in the complex plane]{\includegraphics[scale=0.7] {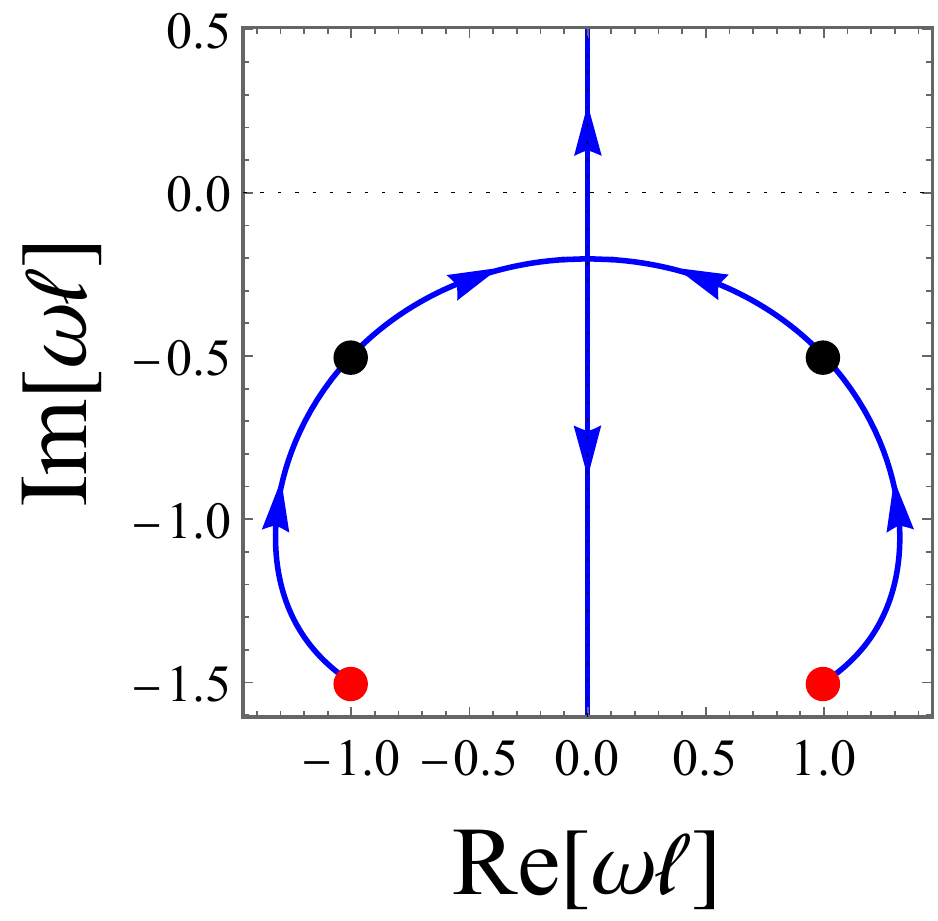}\label{QNF}}
\subfigure[$\textrm{Im}(\omega)$ and~$-\kappa$]{\includegraphics[scale=0.7] {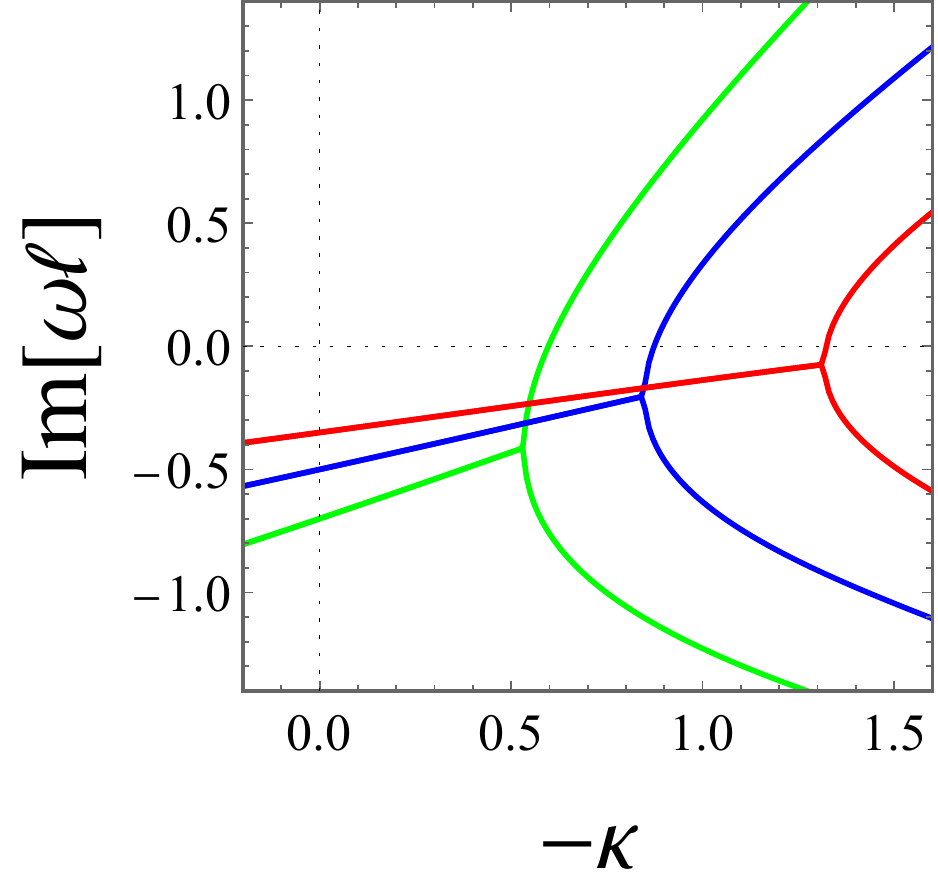}\label{ImQNF}}
\caption{({\it left panel}) Flow of the QNM frequencies with respect to~$\kappa$ for~$m=1, \mu^2\ell^2=-0.75,r_H/\ell=1.0$. The arrows denote the direction to which $-\kappa$ increases. The red and black points represent the QNM frequencies for the Dirichlet condition case in Eq.~\eqref{DBC} and the Neumann condition case in Eq.~\eqref{NBC}, respectively. ({\it right panel}) Relation between the imaginary part of the QNM frequencies and~$-\kappa$ for~$m=1, \mu^2\ell^2=-0.75$, and~$r_H/\ell=1.4,1.0,0.7$, which are denoted by the green, blue, and red lines, respectively.} 
\end{figure}

The QNM frequencies are determined by~$A_{\rm II}/ A_{\rm I}=\kappa$. It is hard, in general, to obtain their analytic expressions except for~$\kappa=\infty, 0$. We numerically solve the 
algebraic equation~$A_{\rm II}/ A_{\rm I}=\kappa$, where $A_{\rm I}$ and $A_{\rm II}$ are functions of $\omega$ and are given by Eqs.~\eqref{Am} and~\eqref{Ap},  respectively.  Figure~\ref{QNF} presents the flow of the QNM frequencies of the fundamental mode with~$m=1, \mu^2\ell^2=-0.75, r_{H}/\ell=1$  in the complex $\omega$-plane when we continuously increase $-\kappa$ in the range of $-\kappa\in[-1000, 3]$. At the beginning~$-\kappa=-1000$, the QNM frequencies are in good agreement with the analytic expressions with the Dirichlet boundary condition, i.e., Eq.~\eqref{DBC}. As $-\kappa$ increases, $|{\rm Im}[\omega]|$ decreases, while $|{\rm Re}[\omega]|$ increases up to $-\kappa\simeq -2$ and turn to the decrease. At $-\kappa=0$, the QNM frequencies are in good agreement with the analytic expressions with the Neumann boundary condition, i.e., Eq.~\eqref{NBC}. As $-\kappa$ further increases, the trajectories approach the imaginary axis with~${\rm Im}[\omega]<0$, and intersect the imaginary axis at~$-\kappa=-\kappa_c\simeq0.84$. Finally, the trajectory splits into two on the imaginary axis, one moving in the positive direction on the axis and the other moving in the negative direction. The dynamics is linearly stable for~$-\kappa\le-\kappa_{c}$. We notice that there is a special value~$-\kappa=-\kappa_0(>-\kappa_c)$ such that $\omega=0$ and below which exponentially growing modes exist. In other words, for $-\kappa>-\kappa_0$, the dynamics is linearly unstable due to the boundary condition.\footnote{This kind of instability has been investigated in~\cite{Dappiaggi:2017pbe} for the rotating BTZ black hole case and in~\cite{Katagiri:2020mvm} for the Reissner-Nordstr\"om-AdS-black hole case. We note that in pure AdS, there is a condition~$\kappa\ge\kappa_0$ for which no exponentially growing modes appear~\cite{Ishibashi:2004wx,Katagiri:2020mvm}. In the BTZ spacetime, the critical value~$\kappa_0$ differs from the pure AdS case.}

Figure~\ref{ImQNF} shows the relation between the imaginary part of the QNM frequencies of the fundamental mode and~$\kappa$. We set~$m=1, \mu^2\ell^2=-0.75$, and~$r_H/\ell=1.4,1.0,0.7$, which are denoted by the green, blue, and red lines, respectively. Each line bifurcates into two at $\kappa=\kappa_c$. It can also be seen that ${\rm Im}[\omega]$ vanishes at $-\kappa=-\kappa_0\simeq0.53, 0.84, 1.29~$ for~$r_H/\ell=1.4,1.0,0.7$, respectively, and unstable modes exist for~$-\kappa>-\kappa_{0}$. 

\subsection{$\mu^2\ell^2=-1$ case}
\begin{figure}[htbp]
\centering
\subfigure[QNM frequencies in the complex plane]{\includegraphics[scale=0.7] {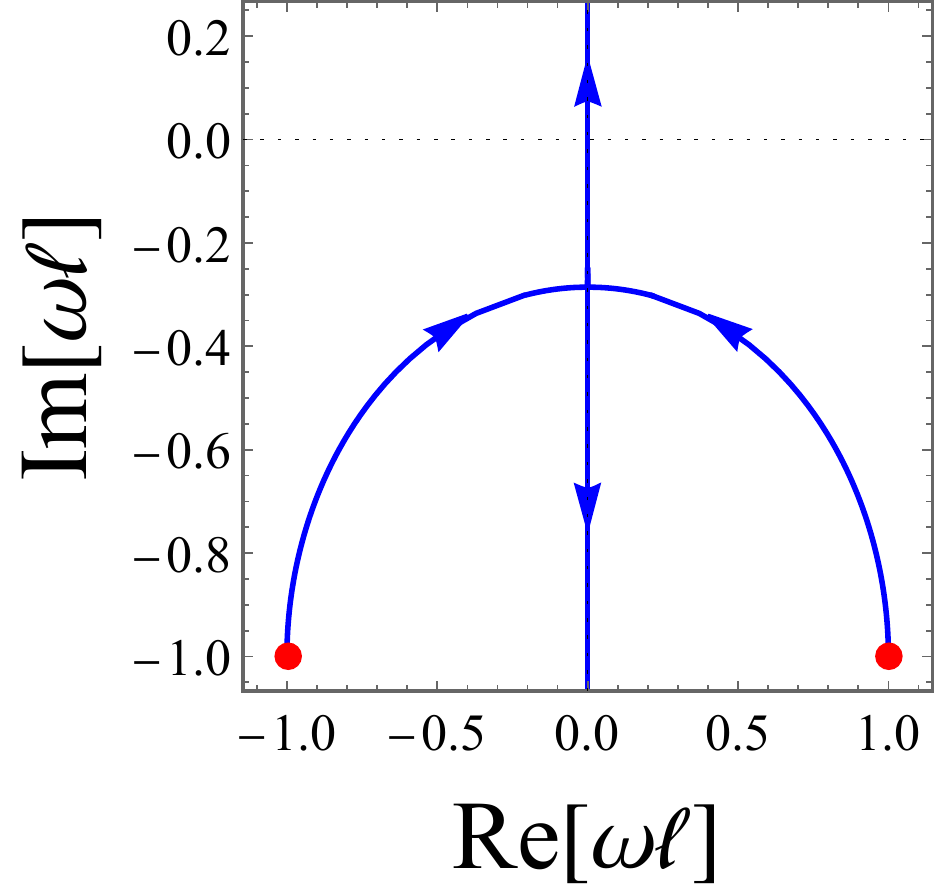}\label{QNFBF}}
\subfigure[$\textrm{Im}(\omega)$ and~$\kappa_{{\rm BF}}$]{\includegraphics[scale=0.7] {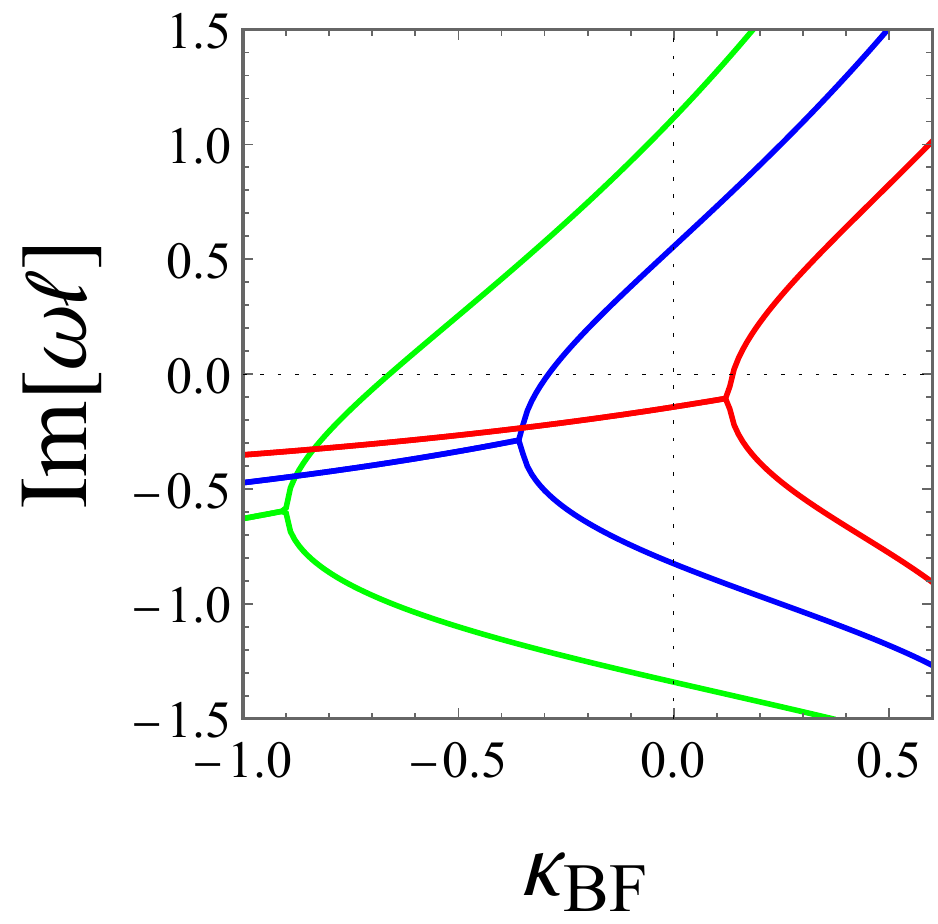}\label{ImQNFBF}}
\caption{({\it left panel})  Flow of the QNM frequencies with respect to~$\kappa_{{\rm BF}}$ for~$m=1, \mu^2\ell^2=-1,r_H/\ell=1.0$. The arrows denote the direction to which $\kappa_{\rm BF}$ increases. The red points represent the QNM frequencies for the Dirichlet-Neumann boundary condition case in Eq.~\eqref{DNBCBF}. ({\it right panel}) Relation between the imaginary part of the QNM frequencies and~$\kappa_{{\rm BF}}$ for~$m=1, \mu^2\ell^2=-1$, and~$r_H/\ell=1.4,1.0,0.7$, which are denoted by the green, blue, and red lines, respectively.} 
\end{figure}

In this case, the Robin boundary condition is~$A_{\rm II, BF}/A_{\rm I, BF}=1/\kappa_{{\rm BF}}$ ($\kappa_{{\rm BF}}\in\mathbb{R}$), where $A_{\rm I, BF}$ and $A_{\rm II, BF}$ are given in Eq.~\eqref{tApAm}. 
We numerically solve~$A_{\rm II, BF}/A_{\rm I, BF}=1/\kappa_{{\rm BF}}$ for $\omega$. Figure~\ref{QNFBF} demonstrates the flow of the QNM frequencies of the fundamental mode with~$m=1, \mu^2\ell^2=-1,r_{H}/\ell=1$ when we increases $\kappa_{{\rm BF}}$ in the range of $\kappa_{{\rm BF}}\in[-1000,1]$. At the beginning $\kappa=-1000$, the QNM frequency is in good agreement with the analytic expression~\eqref{DNBCBF}. As~$\kappa_{{\rm BF}}$ increases, the QNM frequencies approach the imaginary axis with~${\rm Im}[\omega]<0$, and intersect the imaginary axis at~$\kappa_{{\rm BF}}=\kappa_{{\rm BF},c}\simeq-0.38$. Finally, the trajectory splits into two on the imaginary axis, one and the other moving in the positive and negative direction on the axis, respectively. The dynamics is stable for~$\kappa_{{\rm BF}}\le\kappa_{{\rm BF},c}$. There exists a special value $\kappa_{{\rm BF}}=\kappa_{{\rm BF},0} (>\kappa_{{\rm BF},c})$ such that $\omega=0$ and beyond which exponentially growing modes exist.\footnote{In the pure AdS case, there is a condition~$\kappa_{{\rm BF}}\le\kappa_{{\rm BF},0}$ beyond which exponentially growing modes appear~\cite{Ishibashi:2004wx}.
}

Figures~\ref{ImQNFBF} shows that the relation between the imaginary part of the QNM frequencies of the fundamental mode and~$\kappa_{{\rm BF}}$.  We set~$m=1, \mu^2\ell^2=-0.75$, and~$r_H/\ell=1.4,1.0,0.7$ which are denoted by the green, blue, and red lines, respectively. Each line bifurcates into two at~$\kappa_{{\rm BF}}=\kappa_{{\rm BF},c}$. It can also be seen that ${\rm Im}[\omega]$ vanishes at $\kappa_{{\rm BF},0}\simeq -0.92,-0.38,0.84$ for~$r_H/\ell=1.4,1.0,0.7$, respectively, and unstable modes exist for~$\kappa_{{\rm BF}}>\kappa_{{\rm BF},0}$.

\section{Asymptotic behaviors of $D_{2,k}\Phi$ and $D_{3,k}\Phi$}
\label{Appendix:D2D3cases}
We discuss the asymptotic behaviors of $D_{2,k}\Phi$ and $D_{3,k}\Phi$ near the horizon and infinity.
As mentioned in Sec.~\ref{Sec:MassladderandQNMBC}, because the mass ladder operators $D_{2,k}$ and $D_{3,k}$ are not globally regular,
$D_{2,k}\Phi$ and $D_{3,k}\Phi$ are also not globally regular even if $\Phi$ is a globally regular solution.
This can be seen in the factor~$e^{\pm r_H\varphi/\ell}$, i.e. 
the mapped solutions are not globally smooth in $\varphi$ direction, in the following discussion.
Thus, the mass ladder operators $D_{2,k_\pm}$ and $D_{3,k_\pm}$ fail to generate a QNM from the QNM solution $\Phi$. 
However, as is stated in Sec.~\ref{Sec:multipleactionsofD2D3}, the multiple actions with appropriate combinations of $D_{2,k_\pm}$ and $D_{3,k_\pm}$ are regular operators and they can generate a QNM.

\subsection{At the horizon}
We consider the asymptotic behavior of a mapped solution near the horizon. Acting the mass ladder operators~$D_{2,k_\pm}$ and~$D_{3,k_\pm}$ on the exact solution~\eqref{PhiQNM}, the asymptotic behaviors at~$r\to r_H$ are calculated to
\begin{equation}
\begin{split}
D_{2,k_\pm}\Phi =& -2^{-i \frac{\omega\ell^2}{2r_H}}A\frac{i\omega \ell^2+k_\pm r_H-im \ell}{r_H\ell}\left(\frac{r-r_H}{r_H}\right)^{-i\frac{\ell^2}{2r_H}\omega }\left[1+\mathcal{O}(r-r_H)\right]e^{-i\omega  t+i\left(m-i\frac{r_H}{\ell}\right)\varphi},\\
D_{3,k_\pm}\Phi =&2^{-i \frac{\omega\ell^2}{2r_H}}A\frac{i\omega \ell^2+k_\pm r_H+im \ell}{r_H\ell}\left(\frac{r-r_H}{r_H}\right)^{-i\frac{\ell^2}{2r_H}\omega }\left[1+\mathcal{O}(r-r_H)\right]e^{-i\omega t+i\left(m+i\frac{r_H}{\ell}\right)\varphi}.
\end{split}
\end{equation}

\subsection{At infinity: Originally from the QNM with the Dirichlet boundary condition}
\label{Appendix:DirichletBC}
We act the mass ladder operators on $\Phi$, which satisfies the Dirichlet boundary condition. Acting the mass ladder operators~$D_{2,k_+}$ and~$D_{3,k_+}$ on the exact solution~\eqref{PhiQNM} with the Dirichlet boundary condition, the asymptotic behaviors at infinity are
\begin{equation}
\begin{split}
\label{DkrphiD23}
D_{2,k_+}\Phi =&-
\frac{\left(\omega \ell^2- r_H(2+\nu)-im\ell\right)\left(\omega \ell^2+ r_H(2+\nu)+im\ell\right)}{2r_H^2\ell(2+\nu)}A_{\rm II}\left(\frac{r_H}{r}\right)^{3+\nu}\\
&\times\left[1+\mathcal{O}(1/r^2)\right]e^{-i\omega  t+i\left(m-i\frac{r_H}{\ell}\right)\varphi},\\
D_{3,k_+}\Phi =&-\frac{\left(\omega \ell^2+ir_H(2+\nu)+ m\ell\right)\left(\omega \ell^2-i r_H(2+\nu)-m\ell\right)}{2r_H^2\ell(2+\nu)}A_{\rm II}\left(\frac{r_H}{r}\right)^{3+\nu}\\
&\times\left[1+\mathcal{O}(1/r^2)\right]e^{-i\omega t+i\left(m+i\frac{r_H}{\ell}\right)\varphi}.
\end{split}
\end{equation}
Acting the mass ladder operators~$D_{2,k_-}$ and~$D_{3,k_-}$ on the exact solution~\eqref{PhiQNM} with the Dirichlet boundary condition, the asymptotic behaviors at infinity are
\begin{equation}
\begin{split}
D_{2,k_-}\Phi &=
-\frac{2(1+\nu)}{\ell}A_{\rm II}\left(\frac{r_H}{r}\right)^{1+\nu}\left[1+\mathcal{O}(1/r^2)\right]e^{-i\omega t+i\left(m-i\frac{r_H}{\ell}\right)\varphi},\\
D_{3,k_-}\Phi &=\frac{2(1+\nu)}{\ell}A_{\rm II}\left(\frac{r_H}{r}\right)^{1+\nu}\left[1+\mathcal{O}(1/r^2)\right]e^{-i\omega  t+i\left(m+i\frac{r_H}{\ell}\right)\varphi}.
\end{split}
\end{equation}

\subsection{At infinity: Originally from the QNM with the Neumann boundary condition}
\label{Appendix:NeumannBC}
We act the mass ladder operators on $\Phi$, which satisfies the Neumann boundary condition. Acting the mass ladder operators~$D_{2,k_+}$ and~$D_{3,k_+}$ on the exact solution~\eqref{PhiQNM} with the Neumann boundary condition, the asymptotic behaviors at infinity are
\begin{equation}
\begin{split}
D_{2,k_+}\Phi &=\frac{2(1+\nu)}{\ell}A_{\rm I}\left(\frac{r_H}{r}\right)^{-1-\nu}\left[1+\mathcal{O}(1/r^2)\right]e^{-i\omega  t+i\left(m-i\frac{r_H}{\ell}\right)\varphi},\\
D_{3,k_+}\Phi &=-\frac{2(1+\nu)}{\ell}A_{\rm I}\left(\frac{r_H}{r}\right)^{-1-\nu}\left[	1+\mathcal{O}(1/r^2)\right]e^{-i\omega  t+i\left(m+i\frac{r_H}{\ell}\right)\varphi}.
\end{split}
\end{equation}
Acting the mass ladder operators~$D_{2,k_-}$ and~$D_{3,k_-}$ on the exact solution~\eqref{PhiQNM} with the Neumann boundary condition, the asymptotic behaviors at infinity are
\begin{equation}
\begin{split}
D_{2,k_-}\Phi &=\frac{\left(i\omega\ell^2+r_H\nu -i m\ell \right)\left(i\omega\ell^2-r_H\nu +i m\ell \right)}{2r_H^2\ell\nu}A_{\rm I}\left(\frac{r_H}{r}\right)^{1-\nu}\left[1+\mathcal{O}(1/r^2)\right]e^{-i\omega  t+i\left(m-i\frac{r_H}{\ell}\right)\varphi},\\
D_{3,k_-}\Phi &=-
\frac{\left(i\omega\ell^2+r_H\nu +i m\ell \right)\left(i\omega\ell^2-r_H\nu -i m\ell \right)}{2r_H^2\ell\nu}A_{\rm I}\left(\frac{r_H}{r}\right)^{1-\nu}\left[1+\mathcal{O}(1/r^2)\right]e^{-i\omega  t+i\left(m+i\frac{r_H}{\ell}\right)\varphi}.
\end{split}
\end{equation}

\section{Mass ladder operators on the QNMs with the Robin boundary condition}
\label{Appendix:otherBCcase}
Here, we act the mass ladder operators~$D_{i,k_\pm}$ on the QNM with the Robin boundary condition. The asymptotic behaviors of the field with $\mu^2\ell^2=\mu(\nu+2)$ at infinity are
\begin{align}
\Phi|_{r\simeq \infty}^{(\nu>-1)}&=
A_{\rm I}\left[\left(\frac{r_H}{r}\right)^{-\nu}\left(1+\mathcal{O}(1/r^2)\right)
+\kappa\left(\frac{r_H}{r}\right)^{2+\nu}\left(1+\mathcal{O}(1/r^2)\right)\right]e^{-i\omega  t+im\varphi},
\label{asymptoticbehaviornearR1}
\\
\Phi|_{r\simeq \infty}^{(\nu =-1)}&= 
A_{\rm II, BF}\left[\left(\frac{r_H}{r}\right)\ln\left(\frac{r_H}{r}\right)+\kappa_{{\rm BF}}\frac{r_H}{r}+\mathcal{O}(1/r^3)\right]e^{-i\omega  t+im\varphi}.
\label{asymptoticbehaviornearR2}
\end{align}
In the derivation of Eq.~\eqref{asymptoticbehaviornearR1}, for simplicity, we have assumed that $\nu$ is not an integer; thus, the expression of the QNM has no logarithmic term.\footnote{If $\nu$ is an integer, in general, the expression of the QNM has logarithmic terms. Then, the asymptotic behaviors include the contribution of $\Xi$ in Eq.~\eqref{functionXi}.}

\subsection{Asymptotic behaviors at infinity: $\nu>-1$ case}
We act the mass ladder operators on~$\Phi $, which satisfies the Robin boundary condition. Acting the mass ladder operators~$D_{i,k_+}$ on the exact solution~\eqref{PhiQNM} with the Robin boundary condition, the asymptotic behaviors at infinity are
\begin{equation}
\begin{split}
\label{DkrPhiRobin}
D_{0,k_+}\Phi =&-\frac{2(1+\nu)}{\ell^2}A_{\rm I}\left[\left(\frac{r_H}{r}\right)^{-1-\nu}\left(1+\mathcal{O}(1/r^2)\right)\right.\\
&\left.-\frac{\left((2+\nu)r_H-i\omega \ell^2\right)^2+m^2\ell^2}{4r_H^2\left(1+\nu\right)(2+\nu)}\kappa\left(\frac{r_H}{r}\right)^{3+\nu}\left(1+\mathcal{O}(1/r^2)\right)\right]e^{-i\left(\omega +i\frac{r_H}{\ell^2}\right)t+im\varphi},\\
D_{1,k_+}\Phi =&\frac{2(1+\nu)}{\ell^2}A_{\rm I}\left[\left(\frac{r_H}{r}\right)^{-1-\nu}\left(1+\mathcal{O}(1/r^2)\right)\right.\\
&\left.-\frac{\left(( 2+\nu)r_H+i\omega \ell^2\right)^2+m^2\ell^2}{4r_H^2\left(1+\nu\right)(2+\nu)}\kappa\left(\frac{r_H}{r}\right)^{3+\nu}\left(1+\mathcal{O}(1/r^2)\right)\right]e^{-i\left(\omega -i\frac{r_H}{\ell^2}\right)t+im\varphi},\\
D_{2,k_+}\Phi=&\frac{2(1+\nu)}{\ell}A_{\rm I}\left[\left(\frac{r_H}{r}\right)^{-1-\nu}\left(1+\mathcal{O}(1/r^2)\right)\right.\\
&\left.-\frac{\left(( 2+\nu)r_H+im\ell\right)^2+\omega ^2\ell^4}{4r_H^2\left(1+\nu\right)(2+\nu)}\kappa\left(\frac{r_H}{r}\right)^{3+\nu}\left(1+\mathcal{O}(1/r^2)\right)\right]e^{-i\omega t+i\left(m-i\frac{r_H}{\ell}\right)\varphi},\\
D_{3,k_+}\Phi =&-\frac{2(1+\nu)}{\ell}A_{\rm I}\left[\left(\frac{r_H}{r}\right)^{-1-\nu}\left(1+\mathcal{O}(1/r^2)\right)\right.\\
&\left.-\frac{\left(( 2+\nu)r_H-im\ell\right)^2+\omega ^2\ell^4}{4r_H^2\left(1+\nu\right)(2+\nu)}\kappa\left(\frac{r_H}{r}\right)^{3+\nu}\left(1+\mathcal{O}(1/r^2)\right)\right]e^{-i\omega t+i\left(m+i\frac{r_H}{\ell}\right)\varphi}.
\end{split}
\end{equation}
Note that $D_{2,k_+}\Phi$ and $D_{3,k_+}\Phi$ are not globally regular due to the factor~$e^{\pm r_H\varphi/\ell}$. It can be seen for $D_{0,k_+}\Phi$ and $D_{1,k_+}\Phi$ that comparing with Eq.~\eqref{asymptoticbehaviornearR1}, those are the asymptotic forms of a solution with $\mu^2\ell^2=(\nu+1)(\nu+3)$, which satisfies the Robin boundary condition. We notice that defining the Robin boundary condition parameter for the mapped solutions, $\tilde{\kappa}_r$, in the same manner as $\kappa$, that is different from $\kappa$, i.e., 
\begin{equation}
\begin{split}
\label{tildekappar}
\kappa\to\tilde{\kappa}_r=\begin{cases}&-\dfrac{\left(( 2+\nu)r_H-i\omega \ell^2\right)^2+m^2\ell^2}{4r_H^2\left(1+\nu\right)(2+\nu)}\kappa,~~~~~~{\rm for}~~D_{0,k_+},\\
&-\dfrac{\left(( 2+\nu)r_H+i\omega \ell^2\right)^2+m^2\ell^2}{4r_H^2\left(1+\nu\right)(2+\nu)}\kappa,~~~~~~{\rm for}~~D_{1,k_+}.
\end{cases}
\end{split}
\end{equation}
Thus, for $\nu>-1$, the mass ladder operators~$D_{0,k_+}$ and $D_{1,k_+}$ keep the Robin boundary condition but change the Robin boundary condition parameter. In general, $\tilde{\kappa}_r$ is complex even if the original $\kappa$ is real. To our knowledge, the physical interpretation of the solution of which the Robin parameter is complex has not been known. 
As an interesting case, 
when the original frequency~$\omega$ is purely imaginary,   
$\tilde{\kappa}_r$ is real. As can be seen in Figure~\ref{QNF} in Appendix~\ref{appendix:QNMfrequencies}, the purely imaginary~$\omega$ indeed exists. 

Acting the mass ladder operators~$D_{i,k_-}$ on the exact solution~\eqref{PhiQNM} with the Robin boundary condition, the asymptotic behaviors at infinity are
\begin{equation}
\begin{split}
\label{DklPhiRobin}
D_{0,k_-}\Phi =&\frac{2(1+\nu)}{\ell^2}A_{\rm I}\left[\kappa\left(\frac{r_H}{r}\right)^{1+\nu}\left(1+\mathcal{O}(1/r^2)\right)\right.\\
&\left.~~~~~~~~~~~~~~~~-\frac{(\nu r_H+i\omega \ell^2)^2+m^2\ell^2}{4r_H^2\nu(\nu+1)}\left(\frac{r_H}{r}\right)^{1-\nu}\left(1+\mathcal{O}(1/r^2)\right)\right]e^{-i\left(\omega +i\frac{r_H}{\ell^2}\right)t+im\varphi},\\
D_{1,k_-}\Phi =&-\frac{2(1+\nu)}{\ell^2}A_{\rm I}\left[\kappa\left(\frac{r_H}{r}\right)^{1+\nu}\left(1+\mathcal{O}(1/r^2)\right)\right.\\
&\left.~~~~~~~~~~~~~~~~~~~-\frac{\left(\nu r_H -i\omega \ell^2\right)^2+m^2\ell^2}{4r_H^2\nu(\nu+1)}\left(\frac{r_H}{r}\right)^{1-\nu}\left(1+\mathcal{O}(1/r^2)\right)\right]e^{-i\left(\omega -i\frac{r_H}{\ell^2}\right)t+im\varphi},\\
D_{2,k_-}\Phi=&-\frac{2(1+\nu)}{\ell}A_{\rm I}\left[\kappa\left(\frac{r_H}{r}\right)^{1+\nu}\left(1+\mathcal{O}(1/r^2)\right)\right.\\
&\left.~~~~~~~~~~~~~~~~~~~+\frac{\left(\nu r_H -i m\ell\right)^2+\omega ^2\ell^4}{4r_H^2\nu(\nu+1)}\left(\frac{r_H}{r}\right)^{1-\nu}\left(1+\mathcal{O}(1/r^2)\right)\right]e^{-i\omega t+i\left(m-i\frac{r_H}{\ell}\right)\varphi},\\
D_{3,k_-}\Phi =&\frac{2(1+\nu)}{\ell}A_{\rm I}\left[\kappa\left(\frac{r_H}{r}\right)^{1+\nu}\left(1+\mathcal{O}(1/r^2)\right)\right.\\
&\left.~~~~~~~~~~~~~~~+\frac{\left(\nu r_H +i m\ell\right)^2+\omega ^2\ell^4}{4r_H^2\nu(\nu+1)}\left(\frac{r_H}{r}\right)^{1-\nu}\left(1+\mathcal{O}(1/r^2)\right)\right]e^{-i\omega t+i\left(m+i\frac{r_H}{\ell}\right)\varphi}.
\end{split}
\end{equation}
Note that $D_{2,k_-}\Phi$ and $D_{3,k_-}\Phi$ are not globally regular due to the factor~$e^{\pm r_H\varphi/\ell}$. It can be seen for $D_{0,k_-}\Phi$ and $D_{1,k_-}\Phi$ that comparing with Eq.~\eqref{asymptoticbehaviornearR1}, those are the asymptotic forms of a solution with $\mu^2\ell^2=(\nu-1)(\nu+1)$, which satisfies the Robin boundary condition. Defining the Robin boundary condition parameter for the mapped solutions, $\tilde{\kappa}_l$, in the same manner as $\kappa$, that is different from $\kappa$, i.e., 
\begin{equation}
\begin{split}
\label{tildekappal}
\kappa\to\tilde{\kappa}_l=\begin{cases}&-\dfrac{(\nu r_H+i\omega \ell^2)^2+m^2\ell^2}{4r_H^2\nu(\nu+1)\kappa },~~{\rm for}~~D_{0,k_-},\\
&-\dfrac{(\nu r_H-i\omega \ell^2)^2+m^2\ell^2}{4r_H^2\nu(\nu+1)\kappa },~~{\rm for}~~D_{1,k_-}.
\end{cases}
\end{split}
\end{equation}
Thus, for $\nu>-1$, the mass ladder operators~$D_{0,k_-}$ and $D_{1,k_-}$ also keep the Robin boundary condition but change the Robin boundary condition parameter. The resulting parameter~$\tilde{\kappa}_l$ is not  necessarily real but is real at least when the original frequency~$\omega$ is purely imaginary~(see  Figure~\ref{QNF}). 

\subsection{Asymptotic behaviors at infinity: $\nu=-1$ case}
We act the mass ladder operators on~$\Phi $, which satisfies the Robin boundary condition for $\nu=-1$.
Acting them on the exact solution~\eqref{PhiQNM} with the Robin boundary condition, the asymptotic behaviors at infinity are
 \begin{equation}
\begin{split}
\label{DPhBFiRobin}
&D_{0,-1}\Phi=\frac{A_{\rm II,BF}}{\ell^2}e^{-i\left(\omega +i\frac{r_H}{\ell^2}\right)t+im\varphi}\left[\left(1+\mathcal{O}(1/r^2)\right)\right.\\
&\left.-\frac{\left(ir_H+m\ell+\omega\ell^2\right)\left(ir_H-m\ell+\omega\ell^2\right)}{2r_H^2}\frac{-1+2\gamma+\psi(1+a)+\psi(1+b)}{2\gamma+\psi(a)+\psi(b)}\kappa_{\rm BF}\left(\frac{r_H}{r}\right)^2\left(\ln r+\mathcal{O}\left(1\right)\right)\right],\\
&D_{1,-1}\Phi=-\frac{A_{\rm II,BF}}{\ell^2}e^{-i\left(\omega -i\frac{r_H}{\ell^2}\right)t+im\varphi}\left[\left(1+\mathcal{O}(1/r^2)\right)\right.\\
&\left.-\frac{\left(ir_H+m\ell-\omega\ell^2\right)\left(ir_H-m\ell-\omega\ell^2\right)}{2r_H^2}\frac{-1+2\gamma+\psi(a)+\psi(b)}{2\gamma+\psi(a)+\psi(b)}\kappa_{\rm BF}\left(\frac{r_H}{r}\right)^2\left(\ln r+\mathcal{O}\left(1\right)\right)\right],\\
&D_{2,-1}\Phi =-\frac{A_{\rm II,BF}}{\ell}e^{-i\omega  t+i\left(m-i\frac{r_H}{\ell}\right)\varphi}\left[\left(1+\mathcal{O}(1/r^2)\right)\right.\\
&\left.-\frac{\left(ir_H-m\ell+\omega\ell^2\right)\left(ir_H-m\ell-\omega\ell^2\right)}{2r_H^2}\frac{-1+2\gamma+\psi(1+a)+\psi(b)}{2\gamma+\psi(a)+\psi(b)}\kappa_{\rm BF}\left(\frac{r_H}{r}\right)^2\left(\ln r+\mathcal{O}\left(1\right)\right)\right],\\
&D_{3,-1}\Phi =\frac{A_{\rm II,BF}}{\ell}e^{-i\omega  t+i\left(m+i\frac{r_H}{\ell}\right)\varphi}\left[\left(1+\mathcal{O}(1/r^2)\right)\right.\\
&\left.-\frac{\left(ir_H+m\ell+\omega\ell^2\right)\left(ir_H+m\ell-\omega\ell^2\right)}{2r_H^2}\frac{-1+2\gamma+\psi(a)+\psi(1+b)}{2\gamma+\psi(a)+\psi(b)}\kappa_{\rm BF}\left(\frac{r_H}{r}\right)^2\left(\ln r+\mathcal{O}\left(1\right)\right)\right],
\end{split}
\end{equation}
where $a$,~$b$, and~$c$ are given in Eq.~\eqref{abc} and $\mu^2\ell^2=-1$. Note that $D_{2,-1}\Phi$ and $D_{3,-1}\Phi$ are not globally regular due to the factor~$e^{\pm r_H\varphi/\ell}$. It can be seen for $D_{0,-1}\Phi$ and $D_{1,-1}\Phi$ that comparing with Eq.~\eqref{asymptoticbehaviornearR1}, those are the asymptotic forms of a solution with $\nu=0$, which satisfies the Robin boundary condition. Defining the Robin boundary condition parameter for the mapped solutions, $\tilde{\kappa}$, that is different from $\kappa$, i.e.,\footnote{Note that $\kappa_{\rm BF}=0$ does not lead to $\tilde{\kappa}=0$, and instead, gives a finite value.} 
\begin{equation}
\begin{split}
\label{tildekappaBF}
\tilde{\kappa}=\begin{cases}&-\dfrac{\left(ir_H+m\ell+\omega\ell^2\right)\left(ir_H-m\ell+\omega\ell^2\right)}{2r_H^2}\dfrac{-1+2\gamma+\psi(1+a)+\psi(1+b)}{2\gamma+\psi(a)+\psi(b)}\kappa_{\rm BF},~{\rm for}~D_{0,-1},\\
&-\dfrac{\left(ir_H+m\ell-\omega\ell^2\right)\left(ir_H-m\ell-\omega\ell^2\right)}{2r_H^2}\dfrac{-1+2\gamma+\psi(a)+\psi(b)}{2\gamma+\psi(a)+\psi(b)}\kappa_{\rm BF},~~{\rm for}~D_{1,-1}.\\
\end{cases}
\end{split}
\end{equation}
Thus, for $\nu=-1$, the mass ladder operators keep the Robin boundary condition but change the Robin boundary condition parameter. As far as we numerically confirm, $\tilde{\kappa}$ in Eq.~\eqref{tildekappaBF} is real when $\omega$ is purely imaginary~(see Figure~\ref{QNFBF} for an illustration of the existence of the purely imaginary~$\omega$).

\section{Coefficients of the asymptotic forms of the mapped solutions}
\label{appendix:Coefficients}
Here, we present the explicit forms of the coefficients of the asymptotic forms of the mapped solutions~$D_{0,k_\pm}\Phi$ and $D_{1,k_\pm}\Phi$ in Eqs.~\eqref{DPsi}-\eqref{DkphiDN}: 
\begin{equation}
\begin{split}
\label{c01kpkm}
c_{0,k_\pm}&=2^{-1-i\frac{\ell^2}{2r_H}\left(\omega+i\frac{r_H}{\ell^2}\right)}\frac{i A}{r_H\ell^2\left(\omega\ell^2+i r_H\right)}\\
&~~\times\left[
r_H^2\left(k_\pm (2+k_\pm)-\nu(2+\nu)\right)+\left(\omega \ell^2+ m \ell-ik_\pm r_H\right)\left(\omega \ell^2-m \ell-ik_\pm r_H\right)\right],\\
c_{1,k_\pm}&=-2^{1-i\frac{\ell^2}{2r_H}\left(\omega-i\frac{r_H}{\ell^2}\right)}\frac{i \omega A}{r_H}.
\end{split}
\end{equation}

\begin{equation}
\begin{split}
\label{cDkp}
c_{0,k_+}^{\rm (D)}&=-\frac{\left(\omega \ell^2+ m\ell+i(2+\nu)r_H\right)\left(\omega \ell^2-m\ell+i(2+\nu)r_H\right)}{2r_H^2\ell^2\left(2+\nu\right)}A_{\rm II},\\
c_{1,k_+}^{\rm (D)}&=\frac{\left(\omega \ell^2+m\ell-i\left(2+\nu\right)r_H\right)\left(\omega \ell^2-m\ell-i\left(2+\nu\right)r_H\right)}{2r_H^2\ell^2\left(2+\nu\right)}A_{\rm II}.
\end{split}
\end{equation}

\begin{equation}
\begin{split}
\label{cDkm}
c_{0,k_-}^{\rm (D)}&=\frac{2\left(1+\nu\right)}{\ell^2}A_{\rm II},\\
c_{1,k_-}^{\rm (D)}&=-\frac{2\left(1+\nu\right)}{\ell^2}A_{\rm II}.
\end{split}
\end{equation}

\begin{equation}
\begin{split}
\label{cNkp}
c_{0,k_+}^{\rm (N)}&=-\frac{2(1+\nu)}{\ell^2}A_{\rm I},\\
c_{1,k_+}^{\rm (N)}&=\frac{2(1+\nu)}{\ell^2}A_{\rm I}.
\end{split}
\end{equation}

\begin{equation}
\begin{split}
\label{cNkm}
c_{0,k_-}^{\rm (N)}&=\frac{\left(\omega \ell^2+m\ell-i\nu r_H\right)\left(\omega \ell^2-m\ell-i\nu r_H\right)}{2r_H^2\ell^2\nu}A_{\rm I},\\
c_{1,k_-}^{\rm (N)}&=-\frac{\left(\omega \ell^2+m\ell+i\nu r_H\right)\left(\omega \ell^2-m\ell+i\nu r_H\right)}{2r_H^2\ell^2\nu}A_{\rm I}.
\end{split}
\end{equation}

\begin{equation}
\begin{split}
\label{cDN}
c_{0,-1}^{\rm (DN)}&=-\frac{\left(\omega\ell^2+m\ell+ir_H\right)\left(\omega\ell^2-m\ell+ir_H\right)}{2r_H^2\ell^2}\frac{-1+2\gamma+\psi(1+a)+\psi(1+b)}{2\gamma+\psi(a)+\psi(b)}A_{\rm I, BF},\\
c_{1,-1}^{\rm (DN)}&=\frac{\left(\omega\ell^2-m\ell-ir_H\right)\left(\omega\ell^2+m\ell-ir_H\right)}{2r_H^2\ell^2}\frac{-1+2\gamma+\psi(a)+\psi(b)}{2\gamma+\psi(a)+\psi(b)}A_{\rm I, BF}.
\end{split}
\end{equation}
Note that $k_+=-2-\nu$ and $k_-=\nu$ in Eq.~\eqref{c01kpkm}. We also note that the coefficients~$c_{0,k_+}$ and $c_{0,k_+}^{\rm (D)}$ vanish for the fundamental modes with the Dirichlet boundary condition; $c_{0,k_-}$ and $c_{0,k_-}^{\rm (N)}$ vanish for the fundamental modes with the Neumann boundary condition; $c_{0,-1}$ and $c_{0,-1}^{\rm (DN)}$ vanish for the fundamental mode with the Dirichlet-Neumann boundary condition.


\begin{thebibliography}{99}


\bibitem{Cardoso:2017qmj}
V.~Cardoso, T.~Houri and M.~Kimura,
Phys. Rev. D \textbf{96} (2017) no.2, 024044

\bibitem{Cardoso:2017egd}
V.~Cardoso, T.~Houri and M.~Kimura,
Class. Quant. Grav. \textbf{35} (2018) no.1, 015011


\bibitem{Katagiri:2021scx}
T.~Katagiri and M.~Kimura,
Phys. Rev. D \textbf{103} (2021) no.6, 064011


\bibitem{Katagiri:2021xig}
T.~Katagiri and M.~Kimura,
Phys. Rev. D \textbf{105} (2022) no.6, 064062


\bibitem{Brito:2015oca}
R.~Brito, V.~Cardoso and P.~Pani,
Lect. Notes Phys. \textbf{906} (2015), pp.1-237

\bibitem{Blandford:1977ds}
R.~D.~Blandford and R.~L.~Znajek,
Mon. Not. Roy. Astron. Soc. \textbf{179} (1977), 433-456


\bibitem{Hinderer:2007mb}
T.~Hinderer,
Astrophys. J. \textbf{677} (2008), 1216-1220

\bibitem{Binnington:2009bb}
T.~Binnington and E.~Poisson,
Phys. Rev. D \textbf{80} (2009), 084018


\bibitem{Vishveshwara:1970zz}
C.~V.~Vishveshwara,
Nature \textbf{227} (1970), 936-938

\bibitem{Giesler:2019uxc}
M.~Giesler, M.~Isi, M.~A.~Scheel and S.~Teukolsky,
Phys. Rev. X \textbf{9} (2019) no.4, 041060


\bibitem{Cardoso:2017soq}
V.~Cardoso, J.~L.~Costa, K.~Destounis, P.~Hintz and A.~Jansen,
Phys. Rev. Lett. \textbf{120} (2018) no.3, 031103


\bibitem{Regge:1957td}
T.~Regge and J.~A.~Wheeler,
Phys. Rev. \textbf{108} (1957), 1063-1069

\bibitem{Berti:2005ys}
E.~Berti, V.~Cardoso and C.~M.~Will,
Phys. Rev. D \textbf{73} (2006), 064030

\bibitem{Hod:1998vk}
S.~Hod,
Phys. Rev. Lett. \textbf{81} (1998), 4293

\bibitem{Horowitz:1999jd}
G.~T.~Horowitz and V.~E.~Hubeny,
Phys. Rev. D \textbf{62} (2000), 024027

\bibitem{Cardoso:2001hn}
V.~Cardoso and J.~P.~S.~Lemos,
Phys. Rev. D \textbf{63} (2001), 124015

\bibitem{Birmingham:2001pj}
D.~Birmingham, I.~Sachs and S.~N.~Solodukhin,
Phys. Rev. Lett. \textbf{88} (2002), 151301

\bibitem{Kodama:2003jz}
H.~Kodama and A.~Ishibashi,
Prog. Theor. Phys. \textbf{110} (2003), 701-722

\bibitem{Ishibashi:2003ap}
A.~Ishibashi and H.~Kodama,
Prog. Theor. Phys. \textbf{110} (2003), 901-919

\bibitem{Kodama:2003kk}
H.~Kodama and A.~Ishibashi,
Prog. Theor. Phys. \textbf{111} (2004), 29-73

\bibitem{Berti:2009kk}
E.~Berti, V.~Cardoso and A.~O.~Starinets,
Class. Quant. Grav. \textbf{26} (2009), 163001

\bibitem{Isi:2019aib}
M.~Isi, M.~Giesler, W.~M.~Farr, M.~A.~Scheel and S.~A.~Teukolsky,
Phys. Rev. Lett. \textbf{123} (2019) no.11, 111102

\bibitem{Banados:1992wn}
M.~Banados, C.~Teitelboim and J.~Zanelli,
Phys. Rev. Lett. \textbf{69} (1992), 1849-1851


\bibitem{Ichinose:1994rg}
I.~Ichinose and Y.~Satoh,
Nucl. Phys. B \textbf{447} (1995), 340-372


\bibitem{Breitenlohner:1982jf} 
  P.~Breitenlohner and D.~Z.~Freedman,
  Annals Phys.\  {\bf 144}, 249 (1982).

\bibitem{Breitenlohner:1982bm} 
  P.~Breitenlohner and D.~Z.~Freedman,
  Phys.\ Lett.\  {\bf 115B}, 197 (1982).



\bibitem{Ferrari:1984ozr}
V.~Ferrari and B.~Mashhoon,
Phys. Rev. Lett. \textbf{52} (1984) no.16, 1361


\bibitem{Ferrari:1984zz}
V.~Ferrari and B.~Mashhoon,
Phys. Rev. D \textbf{30} (1984), 295-304


\bibitem{Zaslavsky:1991ug}
O.~B.~Zaslavsky,
Phys. Rev. D \textbf{43} (1991), 605-608


\bibitem{Hatsuda:2019eoj}
Y.~Hatsuda,
Phys. Rev. D \textbf{101} (2020) no.2, 024008


 
\bibitem{Natsuume:2020snz}
M.~Natsuume and T.~Okamura,
Phys. Rev. D \textbf{103} (2021) no.6, 066017
 
\bibitem{Noda:2019mzd}
S.~Noda, Y.~Nambu, T.~Tsukamoto and M.~Takahashi,
Phys. Rev. D \textbf{101} (2020) no.2, 023003
 

\bibitem{Dias:2019ery}
O.~J.~C.~Dias, H.~S.~Reall and J.~E.~Santos,
JHEP \textbf{12} (2019), 097
 
\bibitem{Aros:2002te}
R.~Aros, C.~Martinez, R.~Troncoso and J.~Zanelli,
Phys. Rev. D \textbf{67} (2003), 044014

 
\bibitem{Abramowitz:1972}
M. Abramowitz and I.A. Stegun, {\it Handbook of Mathematical Functions with Formulas, Graphs, and Mathematical
Tables} (Dover, New York, 1972).



\bibitem{Ishibashi:2004wx}
A.~Ishibashi and R.~M.~Wald,
Class. Quant. Grav. \textbf{21} (2004), 2981-3014


\bibitem{Witten:2001ua}
E.~Witten,
[arXiv:hep-th/0112258 [hep-th]].



\bibitem{Hertog:2004ns}
T.~Hertog and G.~T.~Horowitz,
Phys. Rev. Lett. \textbf{94} (2005), 221301

\bibitem{Dappiaggi:2017pbe}
C.~Dappiaggi, H.~R.~C.~Ferreira and C.~A.~R.~Herdeiro,
Phys. Lett. B \textbf{778} (2018), 146-154


\bibitem{Katagiri:2020mvm}
T.~Katagiri and T.~Harada,
Class. Quant. Grav. \textbf{38} (2021) no.13, 135026




\end{thebibliography}
\end{document}